\documentclass[10pt,aps,prd,showpacs,twocolumn,superscriptaddress,longbibliography,nofootinbib]{revtex4-1}

\usepackage{epsfig}
\usepackage{graphicx}
\usepackage[normalem]{ulem}
\usepackage{amssymb}
\usepackage{amsmath}
\usepackage{amsfonts}
\usepackage{latexsym}
\usepackage{verbatim}
\usepackage{mathtools}
\usepackage{setspace}
\usepackage{adjustbox}
\usepackage{slashed}
\usepackage{tikz}
\usepackage{psfrag}
\usepackage{comment}
\usepackage{array}
\usepackage{amssymb}
\usepackage{amsmath}
\usepackage{amsthm}
\usepackage{graphicx}
\usepackage{float}
\usepackage{amsfonts,wasysym,epsfig,verbatim,subfigure,bm,mathrsfs,lipsum}
\usepackage{colortbl}
\usepackage{ulem}

\newcommand{\ie}{i.e.,~}

\begin{document}

\title{Magnetized discs and photon rings around  Yukawa-like black holes}
	\author{Alejandro Cruz-Osorio}
	\affiliation{Institut fur Theoretische Physik, Goethe Universit\"at Frankfurt, Max-von-Laue-Str.1, 60438 Frankfurt am Main, Germany}
	\author{Sergio Gimeno-Soler}
	\affiliation{Departamento de Astronom\'ia y Astrof\'isica, Universitat de Val\`encia, C/ Dr.  Moliner 50, 46100, Burjassot (Val\`encia), Spain}
	\author{Jos\'e A. Font}
	\affiliation{Departamento de Astronom\'ia y Astrof\'isica, Universitat de Val\`encia, C/ Dr.  Moliner 50, 46100, Burjassot (Val\`encia), Spain}
	\affiliation{Observatori Astron\`omic, Universitat de Val\`encia, C/ Catedr\'atico Jos\'e Beltr\'an 2, 46980, Paterna (Val\`encia), Spain}
	\author{Mariafelicia De Laurentis}
	\affiliation{Dipartimento di Fisica,  Universit\'a
di Napoli {}``Federico II'', Compl. Univ. di
Monte S. Angelo, Edificio G, Via Cinthia, I-80126, Napoli, Italy}
	\affiliation{INFN Sezione di Napoli, Compl.~Univ.~di Monte S.~Angelo, Edificio~G, Via Cinthia, I-80126, Napoli, Italy}
	\author{Sergio Mendoza}
	\affiliation{Instituto de Astronom\'{\i}a, Universidad Nacional Aut\'onoma de M\'exico, AP 70-264, Ciudad de M\'exico 04510, M\'exico}

\begin{abstract}
We present stationary solutions of geometrically thick discs (or tori) endowed with a self-consistent toroidal magnetic field distribution surrounding a nonrotating black hole in an analytical, static, spherically-symmetric $f(R)$-gravity background. These $f(R)$-gravity models introduce a Yukawa-like modification to the Newtonian potential, encoded in a single parameter $\delta$ which controls the strength of the modified potential and whose specific values affect the disc configurations when compared to the general relativistic case. Our models span different magnetic field strengths, from purely hydrodynamical discs to highly magnetized tori. 
The characteristics of the solutions are identified by analyzing the central density, mass, geometrical size, angular size, and the black hole metric deviations from the Schwarzschild space-time. In the general relativistic limit ($\delta=0$) our models reproduce previous results for a Schwarzschild black hole. For small values of the \( \delta \) parameter, corresponding to $\sim 10\%$ deviations from general relativity, we find variations of $\sim 2 \%$ in the event horizon size, a $\sim 5\%$ shift in the location of the inner edge and center of the  disc, while the outer edge increases by $\sim 10\%$. Our analysis for $|\delta|>0.1$, however, reveals notable changes in the black hole space-time solution which have a major impact in the morphological and thermodynamical  properties of the discs. The comparison with general relativity is further investigated by computing the size of the photon ring produced by a source located at infinity. This allows us to place constraints on the parameters of the $f(R)$-gravity model based on the Event Horizon Telescope observations of the size of the light ring in M87 and SgrA$^*$.
\end{abstract}
\maketitle

\section{Introduction}
\label{sec:intro}

Astrophysical systems comprising a rotating black hole surrounded by an accretion thick disc of plasma are recognized as natural end results of highly dynamical events involving compact objects in a  general-relativistic regime. Stellar-origin systems are produced in mergers of compact binaries comprising either a black hole and a neutron star or two neutron stars, as well as in the gravitational collapse of massive stars (``failed" supernovae)~\citep{,Woosley1993,Baiotti2017}.  Mergers of compact binaries have been dramatically disclosed in recent times thanks to the LIGO-Virgo observations of gravitational waves from GW170817 and the scores of multiwavelength electromagnetic observations that followed~\citep{Abbott:2017a,Abbott:2017b}.  
Additionally, black hole-disc systems are used to explain astrophysical phenomenology of supermassive black holes in active galactic nuclei~\citep{Shakura:1973,Rees:1984}. Major observational 
advances of the strong-gravity region of such systems have recently been accomplished through the ground-breaking first image of the M87 black hole by the Event Horizon Telescope \citep{EHT_M87_PaperI}. 

Theoretical models describing the morphology of stationary thick discs around black holes in general relativity were first developed in the seminal papers of~\citet{Fishbone76} and~\citet{Kozlowski78} for isentropic and barotropic discs respectively, assuming a constant distribution of angular momentum.
The modeling has gradually improved through the elapsing decades \citep[see e.g.][for a review]{Abramowicz13}. 
Proposals to construct the initial data for magnetized discs with weak magnetic fields exploring different configurations were put forward and evolved by various authors, e.g.~advection-dominated accretion flows (ADAF)  \citep{Narayan1994,Yuan2014}, standard and normal evolution (SANE) flows with poloidal magnetic field \citep{Narayan2012, Sadowski2013} -- for a comparison between codes for SANE evolutions see \citep{Porth2019} --, and magnetically arrested dominated (MAD) flows \citep{Narayan2003, DeVilliers03, Tchekhovskoy2011, McKinney2012}. Self-consistent solutions for
magnetized thick discs with toroidal distributions of the magnetic field were obtained by~\citet{Komissarov2006a} for constant angular momentum discs. This solution was extended to the nonconstant angular momentum case by \citet{Montero07} (see also \cite{Gimeno-Soler:2017}) and  by \citet{Pimentel2018a, Pimentel2018b} who incorporated magnetic polarization. 
Recently, equilibrium solutions of self-gravitating magnetized discs in general relativity have been reported by \citet{Mach2019}, building
on a procedure introduced by \citet{Shibata07} for unmagnetized tori (see also~\citet{Stergioulas11}). Numerical evolution of those
solutions have been used to study the development of possible dynamical instabilities in the
discs. These studies include the runaway 
instability~\citep{Abramowicz83}, the Papaloizou-Pringle instability~\citep{Papaloizou84}), the magneto-rotational
instability~\citep{Balbus91} as well as the formation of jets and outflows~\citep[see e.g.][]{Hawley91, Font02, DeVilliers03, Gammie03,Rezzolla2003, Daigne04,Zanotti2005, Fragile07, Montero10, Kiuchi11, Korobkin11, McKinney2012, Korobkin13, Wielgus2015, Mewes16, Fragile2017, Bugli2018, Witzany2018, Cruz2020}). 

\begin{table*}[t]
   \caption{Summary of space-time and disc properties- in geometrized units - for our three values of the length scale $\lambda$ and some representative values of  $\delta$. From left to right the columns report the radii of the event horizon $r_{{}_{\rm EH}}$, of the marginally bound orbit $r_{\rm mb}$, of the cusp of the gravitational potential  $r_{\rm cusp}$, of the inner edge of the disc $r_{\rm in}$, of the center of the disc $r_{\rm c}$, and of the outer edge of the disc $r_{\rm out}$, as well as the specific angular momentum $l_{\rm mb}$ at $r_{\rm mb}$, the gap of the potential $\Delta {\cal W}$, the angular velocity $\Omega$,  and the  orbital period $t_{\rm orb}$ at the center of the disc.
   } 
    \label{tab:params}
    \begin{tabular}{lcccccccccc}    
  	  \hline
 
     \textbf{$\delta$} & \textbf{$r_{{}_{\rm EH}}$} & \textbf{$r_{\rm mb}$}  & \textbf{$r_{\rm cusp}$} & \textbf{$r_{\rm in}$} & \textbf{$r_{\rm c}$}& \textbf{$r_{\rm out}$}&\textbf{$l_{\rm mb}$} &\textbf{$\Delta {\cal W}$}&\textbf{$\Omega$}&\textbf{$t_{\rm orb}$} \\
      \hline
      \multicolumn{11}{c}{\boldmath{$\lambda=10$}}\\     
      \hline
      
      $+0.995$& 1.83 & 3.98 & 3.97  & 4.70  & 9.63   & 120.97 & 3.39  & 0.022 & 0.0313 & 32.0 \\
      $+0.500$& 1.97 & 3.98 & 3.98  & 4.74  & 10.32 & 102.83   & 3.88  & 0.039 & 0.0306 & 33.6 \\ 
      $+0.100$& 1.89 & 3.96 & 3.96  & 4.67  & 9.92   & 99.30     & 3.58  & 0.028 & 0.0298 & 32.7 \\
      $+0.000$& 2.00 & 4.00 & 4.00  & 4.67  & 10.47 & 108.52   & 4.00  & 0.043 & 0.0295 & 33.9 \\
      $-0.100$ & 2.04 & 4.03 & 4.03  & 4.80  & 10.63 & 111.28   & 4.15  & 0.048 & 0.0293 & 34.1 \\
      $-0.300$ & 2.17 & 4.16 & 4.15  & 4.95  & 11.08 & 116.95   & 4.61  & 0.062 & 0.0288 & 34.7\\
      $-0.500$ & 2.43 & 4.50 & 4.51  & 5.35  & 11.89 & 107.87   & 5.57  & 0.081 & 0.0282 & 35.5   \\ 
      $-0.800$ & 5.28 & 9.91 & 9.90  & 11.59& 24.03 & 269.63   & 15.50& 0.099 & 0.0165 & 60.7\\
      $-0.995$ &400.0& 800.0& 800.0& 952.0&2094.4&21704.8 & 800.0& 0.043 & 0.0001 & 6777.7\\
      \hline
      \hline
      \multicolumn{11}{c}{\boldmath{$\lambda=60$}}\\       
      \hline
     
      $+0.995$& 1.97 & 4.00 & 4.01 & 4.67 & 10.04 & 69.45 & 3.87 & 0.036 & 0.0314 & 31.9  \\
      $+0.500$  & 1.98 & 4.00 & 4.01 & 4.67 & 10.16 & 76.27 & 3.91 & 0.038 & 0.0308 & 32.4  \\
      $+0.100$  & 1.99 & 4.00 & 3.99 & 4.72 & 10.43 & 111.80 & 3.98 & 0.043 & 0.0297 & 33.7 \\
      $+0.000$  & 2.00 & 4.00 & 4.00 & 4.67 & 10.47 & 108.52 & 4.00 & 0.043 & 0.0295 & 33.9 \\
      $-0.100$   & 2.01 & 4.00 & 3.99 & 4.67 & 10.57 & 120.04 & 4.03 & 0.045 & 0.0291 & 34.3 \\
     $-0.300$  & 2.03 & 4.01 & 4.01 & 4.77 & 10.78 & 130.68 & 4.11 & 0.048 & 0.0283 & 35.3 \\
      $-0.500$ & 2.07 & 4.02 & 4.02 & 4.78 & 11.26 & 182.45 & 4.27 & 0.057 & 0.0267 & 37.5 \\
      $-0.800$ & 2.30 & 4.22 & 4.21 & 5.06 & 13.28 & 262.41 & 5.20 & 0.099 & 0.0215 & 46.4 \\
      $-0.995$ &399.0&800.0& 799.9&952.0&2094.4&21708.8&800.0&0.043 &0.0001& 6777.7\\
      \hline    
      \hline    
      \multicolumn{11}{c}{\boldmath{$\lambda=1000$ }}   \\ 
      \hline
      
      $+0.995$& 2.00 & 4.00 & 4.00 & 4.76 & 10.42 & 98.48   & 3.99 & 0.042 & 0.0297 & 33.6   \\
      $+0.500$& 2.00 & 4.00 & 4.01 & 4.76 & 10.41 & 94.07   & 3.99 & 0.042 & 0.0298 & 33.6   \\
      $+0.100$& 2.00 & 4.00 & 3.99 & 4.76 & 10.48 & 111.68 & 4.00 & 0.043 & 0.0295 & 33.9  \\
      $+0.000$& 2.00 & 4.00 & 4.00 & 4.76 & 10.47 & 108.52 & 4.00 & 0.043 & 0.0295 & 33.9  \\
      $-0.100$ & 2.00 & 4.00 & 4.00 & 4.76 & 10.46 & 104.90 & 4.00 & 0.043 & 0.0296 & 33.8  \\
      $-0.300$ & 2.00 & 4.00 & 3.99 & 4.76 & 10.53 & 123.53 & 4.01 & 0.044 & 0.0293 & 34.2\\
      $-0.500$ & 2.00 & 4.00 & 3.99 & 4.76 & 10.57 & 135.08 & 4.02 & 0.045 & 0.0291 & 34.4  \\
      $-0.800$ & 2.02 & 4.00 & 4.01 & 4.76 & 10.67 & 147.41 & 4.06 & 0.046 & 0.0287 &  34.8\\ 
      $-0.995$ & 3.32 & 5.50 & 5.51 & 6.60 & 25.16 & 3671.75 & 9.80 & 0.277 & 0.0082 & 122.5 \\
      \hline                    
      \hline    
    \end{tabular}
\end{table*}

All of those studies have been performed within the framework of general relativity in which astrophysical black holes are described by the Schwarzschild or Kerr solutions. However, other types of black hole solutions have been obtained in extended theories of gravity. The observational capabilities offered by the Event Horizon Telescope, targeted to measure black hole shadows and associated strong-field lensing patterns from accretion discs around black holes, allows to test the validity of the black hole solutions of general relativity. As a recent example~\citet{Mizuno18} used the parametrized Einstein-Maxwell-dilaton-axion gravity solutions of~\citet{Garcia1995,Konoplya16} to  compare the shadows from a Kerr black hole and a dilaton one, offering a proof of concept for the feasibility of such tests. More recently, the shadow of a boson star -surfaceless black hole mimicker- was also studied in \citet{Olivares2020}, where considering realistic astronomical observing conditions shows that is possible to distinguish between Kerr black holes and nonrotating boson stars.  The dynamics of charged particles around quasi-Schwarzschild and quasi-Kerr black holes, and particle motion around modified black holes, have been recently investigated in~\cite{Lin2015,Narzilloev2021,Narzilloev2019}.
Motivated by those works here we explore the consequences of a simple extended model of gravity that at ``zeroth'' perturbation order reproduces the standard Schwarzschild space-time geometry of general relativity, but greatly differs from it when additional terms are taken into account. This is done using a pure metric $f(R)$ theory of gravity~\citep{capozziello10,capozziello10b,capozziello11,nojiri17,harko18} by the introduction of a Yukawa-like potential following the proposal of~\citet{DeMartino2018yqf,DeLaurentis2018ahr}. As an astrophysical application, in this article  we explore two directions: (1) building stationary solutions of magnetized thick discs with a self-consistent toroidal magnetic field around a Yukawa-like black hole and (2)  computing the photon ring size in both general relativity and in our extended theory of choice. The two solutions are compared with the aim of exploring whether the extended $f(R)$ theory involving  
a Yukawa-like potential can still be valid with current observations. Previous attempts to construct accretion discs around black holes in $f(R)$ theories of gravity can be found in \citet{Perez2013} and  \citet{Alipour2016} for thin and thick discs, respectively. 
  
This paper is organized as follows: In Sec.~\ref{sec:Num} we summarize the problem setup, describing the black hole space-time in $f(R)$ gravity and the procedure to build the disc solution. Stationary models of thick discs varying the space-time parameters and the disc magnetization are presented in Sec.~\ref{sec:Results}. This section also discusses the dependence of the photon ring size on the space-time parameters Finally Sec.~\ref{sec:Sum} summarizes our conclusions. Unless stated otherwise we use geometrized units in which the light speed, Newton's constant, and the mass of the black hole are equal to one, $c=G=M=1$, the Kerr metric has the signature $(-,+,+,+)$, and the $1/4\pi$ factor in the MHD equations is assumed to be one.     

\begin{table*}
    \caption{Maximum values of the rest-mass density $\rho_{\rm max}$ and of the baryon mass of the disc $M_{\rm disc}$ as a function of the magnetization of the plasma $\beta$. The quantities are shown for all values of the YBH space-time parameters $\delta$ and $\lambda$. We assume that the baryon mass is $M_{\rm disc}=0.1 M$ in the purely hydrodynamical case  $(\beta=10^{3}$).}
  \begin{center}
    \label{tab:rho_and_m}
\begin{adjustbox}{max width=0.785\textwidth}
    \begin{tabular}{c|c|c|c|c|c|c|c}    
  	  \hline
      \textbf{$\delta$}  & \textbf{$\beta=10^{3}$} & \textbf{$\beta=10^{2}$}  & \textbf{$\beta=10^{1}$} & \textbf{$\beta=10^{0}$} & \textbf{$\beta=10^{-1}$}& \textbf{$\beta=10^{-2}$}&\textbf{$\beta=10^{-3}$}\\

      \hline
      \hline
      \multicolumn{8}{c}{\boldmath{$\rho_{\rm max}$}}\\ 
      \hline
      \multicolumn{8}{c}{\boldmath{$\lambda=10$}}\\ 
      \hline
      $+0.995$ & $3.32 \times 10^{-5}$ & $3.32 \times 10^{-5}$ & $3.33 \times 10^{-5}$ &$3.57 \times 10^{-5}$ & $4.10 \times 10^{-5}$ & $4.28 \times 10^{-5}$ & $4.31 \times 10^{-5}$ \\
      $+0.500$ & $2.97 \times 10^{-5}$ & $2.97 \times 10^{-5}$ & $2.98 \times 10^{-5}$ &$3.20 \times 10^{-5}$ & $3.78 \times 10^{-5}$ & $3.93 \times 10^{-5}$ & $3.95 \times 10^{-5}$\\
      $+0.100$ & $2.26 \times 10^{-5}$ & $42.26 \times 10^{-5}$ & $2.27 \times 10^{-5}$ &$2.46 \times 10^{-5}$ & $2.94 \times 10^{-5}$ & $3.08 \times 10^{-5}$ & $3.10 \times 10^{-5}$\\
      $+0.000$ & $2.00 \times 10^{-5}$ & $2.00 \times 10^{-5}$ & $2.00 \times 10^{-5}$ & $2.18 \times 10^{-5}$& $2.62 \times 10^{-5}$ & $2.75 \times 10^{-5}$ & $2.77 \times 10^{-5}$\\
      $-0.100$  & $1.83 \times 10^{-5}$ & $1.83 \times 10^{-5}$ & $1.83 \times 10^{-5}$ & $2.00 \times 10^{-5}$& $2.42 \times 10^{-5}$ & $2.55 \times 10^{-5}$ & $2.56 \times 10^{-5}$\\
      $-0.500$  & $1.43 \times 10^{-5}$ & $1.43 \times 10^{-5}$ & $1.43 \times 10^{-5}$ & $1.56 \times 10^{-5}$& $1.92 \times 10^{-5}$ & $2.03 \times 10^{-5}$ & $2.05 \times 10^{-5}$\\
      $-0.995$  &  $2.45 \times 10^{-12}$& $2.45 \times 10^{-12}$ & $2.50 \times 10^{-12}$ & $2.72 \times 10^{-12}$& $3.28 \times 10^{-12}$ & $3.44 \times 10^{-12}$ & $3.46 \times 10^{-12}$\\
                                           
      \hline
      \multicolumn{8}{c}{\boldmath{$\lambda=60$}}\\ 
      \hline
      $+0.995$ & $4.20 \times 10^{-5}$ & $4.20 \times 10^{-5}$ & $4.21 \times 10^{-5}$ & $4.53 \times 10^{-5}$ & $5.34 \times 10^{-5}$ & $5.59 \times 10^{-5}$ & $5.60 \times 10^{-5}$  \\
      $+0.500$ & $3.53 \times 10^{-5}$ & $3.53 \times 10^{-5}$ & $3.54 \times 10^{-5}$ & $3.80 \times 10^{-5}$ & $4.51 \times 10^{-5}$ & $4.74 \times 10^{-5}$ & $4.75 \times 10^{-5}$ \\
      $+0.100$ & $2.02 \times 10^{-5}$ & $2.02 \times 10^{-5}$ & $2.03 \times 10^{-5}$ & $2.20 \times 10^{-5}$ & $2.65 \times 10^{-5}$ & $2.78 \times 10^{-5}$ & $2.08 \times 10^{-5}$ \\
      $+0.000$ & $2.00 \times 10^{-5}$ & $2.00 \times 10^{-5}$ & $2.00 \times 10^{-5}$ & $2.18 \times 10^{-5}$ & $2.62 \times 10^{-5}$ & $2.75 \times 10^{-5}$ & $2.77 \times 10^{-5}$ \\
      $-0.100$  & $1.67 \times 10^{-5}$ & $1.67 \times 10^{-5}$ & $1.68 \times 10^{-5}$ & $1.81 \times 10^{-5}$ & $2.21 \times 10^{-5}$ & $2.32 \times 10^{-5}$ & $2.34 \times 10^{-5}$ \\
      $-0.500$  & $6.00 \times 10^{-6}$ & $6.00 \times 10^{-6}$ & $6.02 \times 10^{-6}$ & $6.62 \times 10^{-6}$ & $8.30 \times 10^{-6}$ & $8.81 \times 10^{-6}$ & $8.88 \times 10^{-6}$ \\
      $-0.995$  & $2.49 \times 10^{-12}$ & $2.49 \times 10^{-12}$ & $2.50 \times 10^{-12}$ & $2.72 \times 10^{-12}$ & $3.28 \times 10^{-12}$ & $3.44 \times 10^{-12}$ & $3.46 \times 10^{-12}$ \\  
                                            
      \hline
      \multicolumn{8}{c}{\boldmath{$\lambda=1000$}}\\
      \hline
      $+0.995$ & $2.25 \times 10^{-5}$ & $2.25 \times 10^{-5}$ & $2.26 \times 10^{-5}$ & $2.46 \times 10^{-5}$ & $2.94 \times 10^{-5}$ & $3.10 \times 10^{-5}$ & $3.12 \times 10^{-5}$  \\
      $+0.500$ & $2.38 \times 10^{-5}$ & $2.38 \times 10^{-5}$ & $2.39 \times 10^{-5}$ & $2.60 \times 10^{-5}$ & $3.10 \times 10^{-5}$ & $3.27 \times 10^{-5}$ & $3.29 \times 10^{-5}$  \\
      $+0.100$ & $1.93 \times 10^{-5}$ & $1.93 \times 10^{-5}$ & $1.94 \times 10^{-5}$ & $2.10 \times 10^{-5}$ & $2.54 \times 10^{-5}$ & $2.66 \times 10^{-5}$ & $2.68 \times 10^{-5}$  \\
      $+0.000$ & $2.00 \times 10^{-5}$ & $2.00 \times 10^{-5}$ & $2.00 \times 10^{-5}$ & $2.18 \times 10^{-5}$ & $2.62 \times 10^{-5}$ & $2.75 \times 10^{-5}$ & $2.77 \times 10^{-5}$ \\
      $-0.100$  & $2.08 \times 10^{-5}$ & $2.08 \times 10^{-5}$ & $2.08 \times 10^{-5}$ & $2.27 \times 10^{-5}$ & $2.73 \times 10^{-5}$ & $2.86 \times 10^{-5}$ & $2.88 \times 10^{-5}$  \\
      $-0.500$  & $1.54 \times 10^{-5}$ & $1.54 \times 10^{-5}$ & $1.54 \times 10^{-5}$ & $1.68 \times 10^{-5}$ & $2.03 \times 10^{-5}$ & $2.14 \times 10^{-5}$ & $2.16 \times 10^{-5}$  \\
      $-0.995$  & $5.66 \times 10^{-10}$ & $5.67 \times 10^{-10}$ & $5.75 \times 10^{-10}$ & $7.69 \times 10^{-10}$ & $1.41 \times 10^{-9}$ & $1.68 \times 10^{-9}$ & $1.72 \times 10^{-9}$  \\            
      \hline
      \hline                                
      \multicolumn{8}{c}{\boldmath{$M_{\rm disc}=\int \sqrt{\gamma}W \rho d^{3}x$}}\\ 
      \hline
      \multicolumn{8}{c}{\boldmath{$\lambda=10$}}\\ 
      \hline
      $+0.995$ & $1.00 \times 10^{-1}$ & $9.72 \times 10^{-2}$ & $7.73 \times 10^{-2}$ & $3.76 \times 10^{-2}$ & $2.48 \times 10^{-2}$ & $2.35 \times 10^{-2}$ & $2.34 \times 10^{-2}$  \\
      $+0.500$ & $1.00 \times 10^{-1}$ & $9.76 \times 10^{-2}$ & $7.97 \times 10^{-2}$ & $4.07 \times 10^{-2}$ & $2.79 \times 10^{-2}$ & $2.62 \times 10^{-2}$ & $2.61 \times 10^{-2}$  \\
      $+0.100$ & $1.00 \times 10^{-1}$ & $9.76 \times 10^{-2}$ & $8.00 \times 10^{-2}$ & $4.14 \times 10^{-2}$ & $2.81 \times 10^{-2}$ & $2.65 \times 10^{-2}$ & $2.64 \times 10^{-2}$  \\
      $+0.000$ & $1.00 \times 10^{-1}$ & $9.76 \times 10^{-2}$ & $7.97 \times 10^{-2}$ & $4.07 \times 10^{-2}$ & $2.76 \times 10^{-2}$ & $2.60 \times 10^{-2}$ & $2.59 \times 10^{-2}$  \\
      $-0.100$  & $1.00 \times 10^{-1}$ & $9.76 \times 10^{-2}$ & $7.97 \times 10^{-2}$ & $4.08 \times 10^{-2}$ & $2.77 \times 10^{-2}$ & $2.62 \times 10^{-2}$ & $2.60 \times 10^{-2}$  \\
      $-0.500$  & $1.00 \times 10^{-1}$ & $9.79 \times 10^{-2}$ & $8.23 \times 10^{-2}$ & $4.53 \times 10^{-2}$ & $3.20 \times 10^{-2}$ & $3.05 \times 10^{-2}$ & $3.03 \times 10^{-2}$  \\
      $-0.995$  & $1.00 \times 10^{-1}$ & $9.76 \times 10^{-2}$ & $7.97 \times 10^{-2}$ & $4.07 \times 10^{-2}$ & $2.76 \times 10^{-2}$ & $2.60 \times 10^{-2}$ & $2.59 \times 10^{-2}$  \\
                        
      \hline
      \multicolumn{8}{c}{\boldmath{$\lambda=60$}}\\ 
      \hline
      $+0.995$ & $1.00 \times 10^{-1}$ & $9.83 \times 10^{-2}$ & $8.51 \times 10^{-2}$ & $5.11 \times 10^{-2}$ & $3.75 \times 10^{-2}$ & $3.59 \times 10^{-2}$ & $3.56 \times 10^{-2}$   \\
      $+0.500$ & $1.00 \times 10^{-1}$ & $9.82 \times 10^{-2}$ & $8.40 \times 10^{-2}$ & $4.86 \times 10^{-2}$ & $3.51 \times 10^{-2}$ & $3.36 \times 10^{-2}$ & $3.33 \times 10^{-2}$   \\
      $+0.100$ & $1.00 \times 10^{-1}$ & $9.75 \times 10^{-2}$ & $7.95 \times 10^{-2}$ & $4.05 \times 10^{-2}$ & $2.74 \times 10^{-2}$ & $2.58 \times 10^{-2}$ & $2.57 \times 10^{-2}$  \\
      $+0.000$ & $1.00 \times 10^{-1}$ & $9.76 \times 10^{-2}$ & $7.97 \times 10^{-2}$ & $4.07 \times 10^{-2}$ & $2.76 \times 10^{-2}$ & $2.60 \times 10^{-2}$ & $2.59 \times 10^{-2}$ \\
      $-0.100$  & $1.00 \times 10^{-1}$ & $9.74 \times 10^{-2}$ & $7.82 \times 10^{-2}$ & $3.81 \times 10^{-2}$ & $2.56 \times 10^{-2}$ & $2.40 \times 10^{-2}$ & $2.39 \times 10^{-2}$  \\ 
      $-0.500$  & $1.00 \times 10^{-1}$ & $9.62 \times 10^{-2}$ & $7.08 \times 10^{-2}$ & $2.78 \times 10^{-2}$ & $1.67 \times 10^{-2}$ & $1.55 \times 10^{-2}$ & $1.54 \times 10^{-2}$  \\
      $-0.995$  & $1.00 \times 10^{-1}$ & $9.76 \times 10^{-2}$ & $7.97 \times 10^{-2}$ & $4.07 \times 10^{-2}$ & $2.76 \times 10^{-2}$ & $2.60 \times 10^{-2}$ & $2.59 \times 10^{-2}$  \\  
                         
      \hline
      \multicolumn{8}{c}{\boldmath{$\lambda=1000$}}\\ 
      \hline
      $+0.995$ & $1.00 \times 10^{-1}$ & $9.77 \times 10^{-2}$ & $8.08 \times 10^{-2}$ & $4.27 \times 10^{-2}$ & $2.93 \times 10^{-2}$ & $2.78 \times 10^{-2}$ & $2.76 \times 10^{-2}$   \\
      $+0.500$ & $1.00 \times 10^{-1}$ & $9.78 \times 10^{-2}$ & $8.14 \times 10^{-2}$ & $4.37 \times 10^{-2}$ & $3.01 \times 10^{-2}$ & $2.86 \times 10^{-2}$ & $2.85 \times 10^{-2}$   \\
      $+0.100$ & $1.00 \times 10^{-1}$ & $9.75 \times 10^{-2}$ & $7.93 \times 10^{-2}$ & $4.00 \times 10^{-2}$ & $2.72 \times 10^{-2}$ & $2.55 \times 10^{-2}$ & $2.54 \times 10^{-2}$   \\
      $+0.000$ & $1.00 \times 10^{-1}$ & $9.76 \times 10^{-2}$ & $7.97 \times 10^{-2}$ & $4.07 \times 10^{-2}$ & $2.76 \times 10^{-2}$ & $2.60 \times 10^{-2}$ & $2.59 \times 10^{-2}$   \\
      $-0.100$  & $1.00 \times 10^{-1}$ & $9.76 \times 10^{-2}$ & $8.01 \times 10^{-2}$ & $4.14 \times 10^{-2}$ &$2.83 \times 10^{-2}$ & $2.66 \times 10^{-2}$ & $2.65 \times 10^{-2} $  \\
      $-0.500$  & $1.00 \times 10^{-1}$ & $9.72 \times 10^{-2}$ & $7.70 \times 10^{-2}$ & $3.65 \times 10^{-2}$ & $2.40 \times 10^{-2}$ & $2.26 \times 10^{-2}$ & $2.25 \times 10^{-2}$  \\
      $-0.995$  & $1.00 \times 10^{-1}$ & $8.17 \times 10^{-2}$ & $2.38 \times 10^{-2}$ & $2.08 \times 10^{-3}$ & $6.69 \times 10^{-4}$ & $5.68 \times 10^{-4}$ & $5.58 \times 10^{-4}$ \\            
      \hline
    \end{tabular}
 \end{adjustbox}
  \end{center}
\end{table*}
\section{Setup}
\label{sec:Num}

The immediate generalization of the Einstein equations is done by allowing the Ricci scalar \( R \) in the gravitational action to be a general analytical function \( f(R) \) \citep[see e.g.][]{Sotiriou2010,capozziello10,capozziello10b,capozziello11,nojiri17,harko18}, with  the matter action written in its usual form~\citep[see e.g.][]{Landau-Lifshitz1,mendoza20}.  The field equations in this pure metric construction are then obtained  by the null variations of the whole action, i.e. the sum of the  gravitational and matter actions, 
with respect to the space-time metric.  The obtained field equations turn out to be fourth-order differential equations for the metric and therefore, finding solutions of a well-posed particular problem constitute a much harder task.  By construction, when \( f(R) = R \), the Einstein field equations are recovered and the differential field equations are of second order in the metric.

\subsection{Spherically symmetric black hole space-time in a pure metric static $f(R)$ model}
\label{sub:spacetime} 

  In this article we consider a static spherically symmetric space-time. 
The field equations are obtained by the a specific choice of an \( f(R) \) 
function.  In order to provide a general scenario, \citet{capozziello07} 
showed that it was possible to find a weak-field limit solution that can be
satisfied for all analytic \( f(R) \) functions (see also \citet{capozziello11} and references therein).  The idea is to expand in a Taylor series
the function \( f(R) \) and keep terms up to order \( 1 / c^2 \) in the
field equations.  The resulting field equations at that order of approximation
have a Yukawa-like potential solution and, as shown by 
\citet{DeLaurentis2018ahr} at that perturbation order, the general solution 
can be written as~\citep[see also][]{demartino14,DeMartino2018yqf}:

\begin{eqnarray}\label{eq:Yukawa}
ds^2=- \left[1+\Phi(r) \right] dt^2 + \left[ 1-\Phi(r) \right] dr^2 + r^2d\tilde{\Omega}^2\,,
\end{eqnarray}
with
\begin{eqnarray}
\Phi(r) &=& -\frac{2 M \left(\delta  e^{-\frac{r}{\lambda}}+1\right)}{r(\delta +1)}\,. 
\end{eqnarray}

This Yukawa-like black hole (YBH) solution constitutes a generalization of the Schwarzschild space-time for all 
analytic \( f(R) \) functions in the post-Newtonian limit~\citep[see e.g.][]{willbook}. In the previous two equations $d\tilde{\Omega}^2\equiv d\theta^2 + \sin^2 \theta d\phi^2$ is the angular
displacement and $\Phi(r)$  is a Yukawa-like potential. 
Mathematically, the 
constant parameters  \(\lambda\) and \( \delta \) are related to the coefficients of the Taylor expansion of the function \( f(R) \) about a fixed \( R_0 \).  In fact \( \lambda :=  \sqrt{ -6 f''_0 / f'_0} \) and \( \delta := f'_0 - 1 \), where \( [\ ]' :=  \mathrm{d} [\ ] / \mathrm{d} R \).  As such, when \( f(R) = R \), the gravitational action becomes the Hilbert action of general relativity and so \( \delta = 0 \) which leads to the Newtonian potential \( \phi(r) = 
- M /r \) for a point mass source and equation~\eqref{eq:Yukawa} converges to the  Schwarzschild exterior solution.

The parameter $\lambda$ is a length scale which can in principle be adjusted depending on the spatial scale of the particular astrophysical system (see below). Moreover, $\delta$ is the parameter of the theory and governs the strength of the Yukawa-like potential (general relativity and therefore Newton's potential is recovered when $\delta = 0$). The event horizon of the YBH is computed in the same way as for the Schwarzschild black hole,  solving the condition $\rm g_{tt}(r)=0$, where the surface of infinite redshift and the event horizon coincide. Since Yukawa's potential has a nonlinear dependence on the radial coordinate we obtain a transcendental equation for $\rm r_{EH}$. Hence, we use a Newton-Raphson root-finder  to compute the event horizon.

For our YBH solution the angular velocity of Keplerian circular orbits around the black hole reads
\begin{eqnarray}
\Omega^2_{\mathrm{K}}(r)=\frac{M\delta e^{-r/\lambda}}{r^{2}\lambda(\delta+1)} - \frac{\Phi}{2r^{2}}\,. 
\end{eqnarray}
For circular orbits, we can write the angular velocity and the specific angular momentum in terms of the nonzero components of the 4-velocity $u^{\mu}$, namely $\Omega = u^{\phi}/u^{t}$ and $l = - u_{\phi}/u_t$. Using this  the expression for the Keplerian specific angular momentum can be written as
\begin{equation} 
l_{\rm K} = \frac{\pm r^{2}}{1+\Phi} \sqrt{\frac{M\delta e^{-r/\lambda}}{r^{2}\lambda(\delta+1)} - \frac{\Phi}{2r^{2}}}.
\label{eq:Kep_lmb}
\end{equation}
We also define the specific bound angular momentum function $l_{\mathrm{b}}(r)$ which corresponds to the specific angular momentum of a marginally bound orbit at a certain radius $r$ and can be written in our case as
\begin{equation}
l^{2}_{\mathrm{b}}(r)= -\frac{r^{2}\Phi}{1+ \Phi}.
\end{equation}
If we consider prograde (retrograde) motion, finding the minimum (maximum) of the function $l_{\mathrm{b}}(r)$ gives the location of the innermost marginally bound circular orbit $r_{\mathrm{mb}}$ and the value of the specific angular momentum there ($l_{\mathrm{b}}(r_{\mathrm{mb}}) = l_{\mathrm{mb}}$). It is also worth noticing that at said point, the Keplerian angular momentum is equal to $l_{\mathrm{mb}}$ as well (i.e.~$l_{\mathrm{K}}(r_{\mathrm{mb}}) = l_{\mathrm{mb}}$).

%

\subsection{Procedure to build equilibrium magnetized thick discs}
\label{sub:InitTori}

We build sequences of equilibrium thick discs endowed with a toroidal magnetic field in YBH space-time following the procedure first presented in \cite{Komissarov2006a} and generalized by \cite{Montero07,Gimeno-Soler:2017}. 
For simplicity we assume that the plasma in the disc obeys a constant distribution of specific angular momentum $l=l_{\rm K}(r_{\rm mb})$ given by the Keplerian angular momentum equation~\eqref{eq:Kep_lmb}, evaluated at $r_{\rm mb}$. The fundamental equation to describe a non-self-gravitating  equilibrium torus around a black hole is obtained by  applying the projection  tensor $h^{\alpha}_{\,\,\beta} = \delta^{\alpha}_{\,\,\beta} + u^{\alpha}u_{\beta}$ to the conservation  law of the energy-momentum tensor \citep{Gimeno-Soler:2017}. This equation reads 
\begin{equation}\label{eq:diff_ver}
\partial_i(\ln |u_t|) - \frac{\Omega \partial_i l}{1-l\Omega} + \frac{\partial_i p}{\rho h} + \frac{\partial_i \left[\mathcal{L}b^2\right]}{2\mathcal{L} \rho h} = 0\,,
\end{equation}
where $i = r, \theta$.
To obtain the previous equation we have assumed that the thermodynamical relationship between the rest-mass density $\rho$ and the thermal pressure $p$ is given by a barotropic equation of state (EoS), $\rho = \rho(p)$. In particular, we choose a polytropic EoS such as $p = K \rho^\Gamma$ and an EoS for the magnetic pressure $p_{\mathrm{m}} \equiv b^2/2$ such as $p_{\mathrm{m}} = K_{\mathrm{m}} \mathcal{L}^{q-1} (\rho h)^q$, where $K$, $K_{m}$, $q$ and $\Gamma$ are constants and $\mathcal{L} \equiv g_{t\phi}^2 - g_{tt}g_{\phi\phi}$. Moreover, $h$ and $b^2$ in Eq.~(\ref{eq:diff_ver}) are the enthalpy and the modulus (squared) of the magnetic field 4-vector. Using this relations we can rewrite Eq.~\eqref{eq:diff_ver} as 
\begin{eqnarray}\label{eq:solution}
&&{\cal W} - {\cal W}_{\mathrm{in}} + \ln \left(1+ \frac{K\Gamma}{\Gamma-1}\rho^{\Gamma-1}\right) + \nonumber \\ 
&&\frac{q K_{\mathrm{m}}}{q-1}\left[\mathcal{L}\left(\rho + \frac{K\Gamma \rho^{\Gamma}}{\Gamma-1}\right)\right]^{q-1} = 0,
\end{eqnarray}
where ${\cal W} = \ln |u_t|$ is the (gravitational plus centrifugal) potential. To solve Eq.~\eqref{eq:solution} we fix $q = \Gamma=4/3$, the density at the center of the disc $\rho_{\rm c}=1$, the specific  angular momentum $l=l_{\rm mb}$ and we fill $80\%$ of the potential gap $\Delta {\cal W}\equiv {\cal W}_{\rm in}-{\cal W}_{\rm cusp}$, where subindices `in' and `cusp' indicate that the potential is calculated at the inner edge of the disc or at the cusp (see below). Therefore, in our models the discs will always be inside their corresponding Roche lobes ($\Delta {\cal W}<0$).
Models are built using a numerical $(r,\theta)$ grid in a domain $r\in [r_{\rm EH},r_{\rm out}]$, whose specific values are reported in Table~\ref{tab:params}. The number of zones in our base grid is $252 \times 256$ in $r$ and $\theta$, respectively.

\section{Results}
\label{sec:Results}

\subsection{YBH parameters}

Before constructing the YBH-disc solutions we explore suitable values of the freely specifiable parameters of the theory, $\lambda$ and $\delta$. This serves the purpose of understanding the intrinsic properties of the space-time and offers the possibility of comparing our findings with the analysis of \cite{DeMartino2018yqf, DeLaurentis2018ahr}. We consider three length scales, $\lambda=10,\, 60,$ and $1000$ in geometrized units. In physical units and for the case of M87 the first case corresponds to the scale of the black hole photon ring shadow, $\lambda_{\rm phys}(10)\sim 3.11\times 10^{-3}\ {\rm pc}\ (23 \mu {\rm as})$, the second one to the size of the inner core of the jet,  $\lambda_{\rm phys}(60)\sim 1.866\times 10^{-2}\ {\rm pc}\ (230 \mu {\rm as})$, and the third case to the large scale jet of M87, $\lambda_{\rm phys}(10^{3})\sim3.11\times 10^{-1}\ {\rm pc}\ (3.8 {\rm mas})$. Similarly, for the galactic center SgrA* the corresponding values are  $\lambda_{\rm phys}(10)\sim1.985\times 10^{-6}\ {\rm pc}\ (50 \mu {\rm as})$, $\lambda_{\rm phys}(60)\sim 1.191\times 10^{-5}\ {\rm pc}\ (300 \mu {\rm as})$, and $\lambda_{\rm phys}(10^{3})\sim 1.985\times 10^{-4}\ {\rm pc}\ (5 {\rm mas})$\footnote{To estimate the length scales in $\mu {\rm as}$ we assume the following black hole masses and distances to the source: $\rm M_{M87}=(6.2\pm0.7)\times 10^{9}\ M_{\odot}$ and $\rm D=16.8 \pm 0.8$ Mpc for M87 \citep{EHT_M87_PaperI}, and $\rm M_{ SgrA*}=(4.148\pm 0.014)\times10^{6}\ M_{\odot}$ and $\rm D=8.178$ Mpc for the galactic center SgrA* \citep{Gravity2019}.}. For each value of $\lambda$ we use 14 values of $\delta$. We focus our attention in the case $\delta<0$ which is where  more noticeable changes with respect to general relativity are observed. Taking into account these considerations, we build $\rm 294$ magnetized accretion discs around YBHs varying the space-time parameters $\delta$ and $\lambda$ and the strength of the toroidal magnetic field at the center of the tori, $\beta_{\rm c}$.

\begin{figure*}
\centering
\includegraphics[width=2.1\columnwidth,height=1.15\columnwidth]{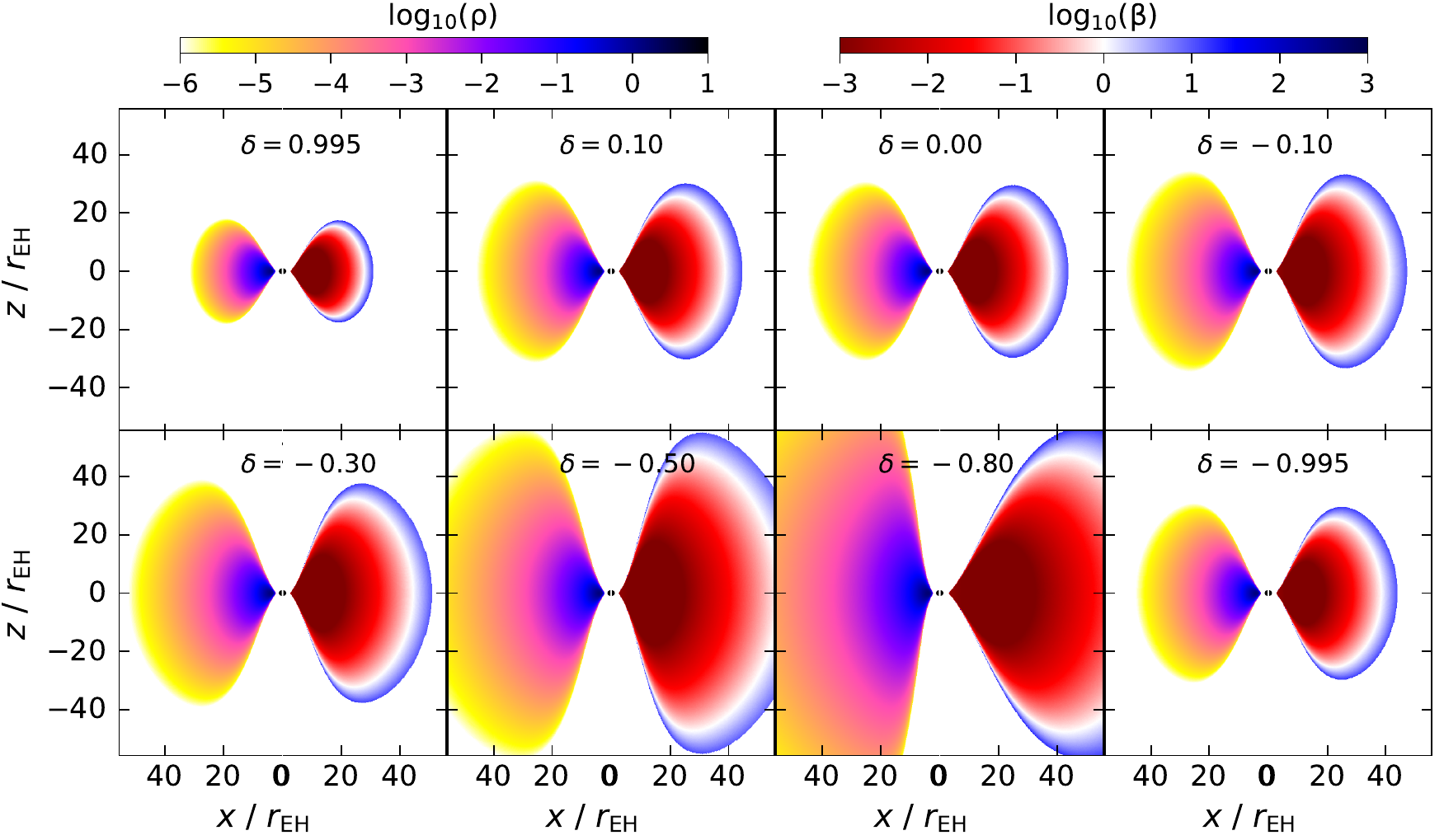}
\caption{Logarithm of the rest-mass density (left half portion of each panel) and magnetization parameter (right half) for an illustrative sample of tori around YBHs with $\lambda=60$ and different values of $\delta$.  All tori are built assuming high magnetization, using a central magnetization parameter of $\beta_c=10^{-3}$. Each box covers a spatial domain $[-50 ~r_{\rm EH}, 50~ r_{\rm EH}]$ in the $x-z$ plane. Note that each solution has a different length scale if expressed in conventional black hole mass $M$ units, due to the dependence of $r_{\rm EH}$ with $\delta$  
(see Table \ref{tab:params} for details). }
\label{fig:2dmorphology}
\end{figure*}

  Note that the choices \( \delta < 0 \) are to be taken with care since they
may produce modifications around a Schwarzschild space-time with \( M < 0 \)
\citep[see e.g][]{townsend}.  For the purpose of this article, we have selected
models for which \( \Phi(r) < 0 \) in order to avoid a negative mass 
Schwarzschild-like spherical solution.

\subsection{Tori geometry}

Table \ref{tab:params} reports the values of selected geometrical quantities of the discs for a subset of representative models (parametrized by $\delta$) for all three values of $\lambda$. We note that those quantities are only related to the space-time and hence do not depend on the magnetization parameter of the discs. Varying parameter $\delta \in [-1,\, 1]$ yields YBHs with different event horizon radii $r_{{}_{\rm EH}}$. We find that the event horizon size increases for negative values of $\delta$ and small values of $\lambda$ reaching  $r_{\rm EH}=400M$ for $\lambda=10$ and $60$. On the other hand, for $\lambda=1000$ the variation of $r_{\rm EH}$ with negative $\delta$ is not too pronounced, staying at $r_{\rm EH}=2$ for most models and reaching $r_{\rm EH}=3.32$ for $\delta=-0.995$. For positive $\delta$ values, the values of $r_{\rm EH}$ we obtain are comparable to the event horizon size of a slowly rotating black hole with $r_{\rm EH}=1.83$ and $1.97$ (for $\lambda=10$ and $60$) corresponding to  Kerr black holes with spin parameters $a=0.558,$ and $0.243$, respectively. Moreover, for positive $\delta$ and $\lambda=1000$ no changes are observed with respect to the event horizon radius of a nonrotating black hole in general relativity. 

Table \ref{tab:params} also reports the radius of the marginally bound orbit $r_{\rm mb}$ and its corresponding specific angular momentum $l_{\rm mb}$. For all $\lambda$, these two quantities show a weak dependence on $\delta$ except for extreme values very close to $\delta=-1$. Additional disc radii reported in Table \ref{tab:params} are the location of the cusp, $r_{\rm cusp}$, the center of the disc, $r_{\rm c}$, and the inner and outer edges of the disc, $r_{\rm in}$, $r_{\rm out}$, respectively. The latter are computed assuming the discs fill $80\%$ of the gap of the potential $\Delta {\cal W}$ (i.e.~of their Roche lobes). The center of the disc is defined as the minimum of ${\cal W}$. The orbital period of the disc, $t_{\rm orb}$, reported in the last column of Table \ref{tab:params} is measured at the center of the disc. It is  found that all characteristic quantities defining the torus size, $r_{\rm cusp}$, $r_{\rm c}$, $r_{\rm in}$, and $r_{\rm out}$, only show a strong dependence on the YBH parameters as $\delta\rightarrow -1$. For all models those quantities increase fairly slowly (or barely increase at all) as $\delta$ goes from positive to negative values, except for $\delta=-0.995$ where the increment is significantly larger.  

The gap of the potential, $\Delta{\cal W}$, also reported in Table \ref{tab:params}, defines the regions where the equilibrium plasma is located around the black hole. In particular, the surface of the magnetized disc is defined as the equipotential surface with ${\cal W}={\cal W}_{\rm in}$ and the solution of the thermodynamical  quantities (see Eq.~\eqref{eq:solution}) also depends on this gap.  In general relativity ($\delta=0$) this value is $\Delta {\cal W}=0.043$, irrespective of $\lambda$. The largest deviations found are  $\Delta {\cal W}=0.112$ (for $ \delta=-0.7),\, 0.176$ (for $\delta=-0.9),$ and $0.277$ (for $\delta=-0.995)$, respectively. 

Figure \ref{fig:2dmorphology} shows the two-dimensional morphology of a representative sample of models. We plot the rest-mass density (left side of each panel) and the  magnetization parameter (right side of each panel) of the tori, the two quantities in logarithmic scale. Results are shown for a YBH space-time with $\lambda=60$ and for different values of $\delta$ and considering a  magnetization parameter at the center of the disc of $\beta_{\rm c}=10^{-3}$. Therefore, the examples shown in Fig.~\ref{fig:2dmorphology} correspond to highly magnetized tori. For clarity in the comparison we rescale the spatial domain $(r \cos\theta,\ r \sin\theta)$ by the event horizon size of each YBH (the specific values are reported in Table \ref{tab:params}) showing a domain $[-50 ~r_{\rm EH}, 50~ r_{\rm EH}]$ in the $x-z$ plane. 
 
As $\delta$ increases from 0 toward $\delta=1$, the size of the disc decreases while $r_{\rm EH}$ is kept essentially constant, slightly changing from 2.0 to 1.97 (see Table \ref{tab:params}). The most noticeable modifications are visible, however, only for the largest values. For $\delta=0.995$ the disc is about 40\% smaller than in general relativity. 
On the other hand, as $\delta$ decreases from 0 toward $\delta=-1$, the size of the discs becomes gradually larger. The largest value in Fig.~\ref{fig:2dmorphology} corresponds to the $\delta=-0.995$ case. Note that the apparent smaller size of this model as compared e.g.~with the $\delta=-0.8$ case is because the domain plotted in the figure is expressed in units of $r_{\rm EH}$ which is 2.3 for the latter and about 400 for the former. Despite these significant modifications in the geometrical size, the distribution of the density and of the magnetization in the torus seem only weakly affected by the changes in $\delta$. For the most negative values of $\delta$ further extended low-density layers are obtained as well as  high-density regions along the symmetry axis of the black hole. Similarly,  the angular thickness of the disc also increases. 

\begin{figure}
\centering
\includegraphics[width=0.95\columnwidth]{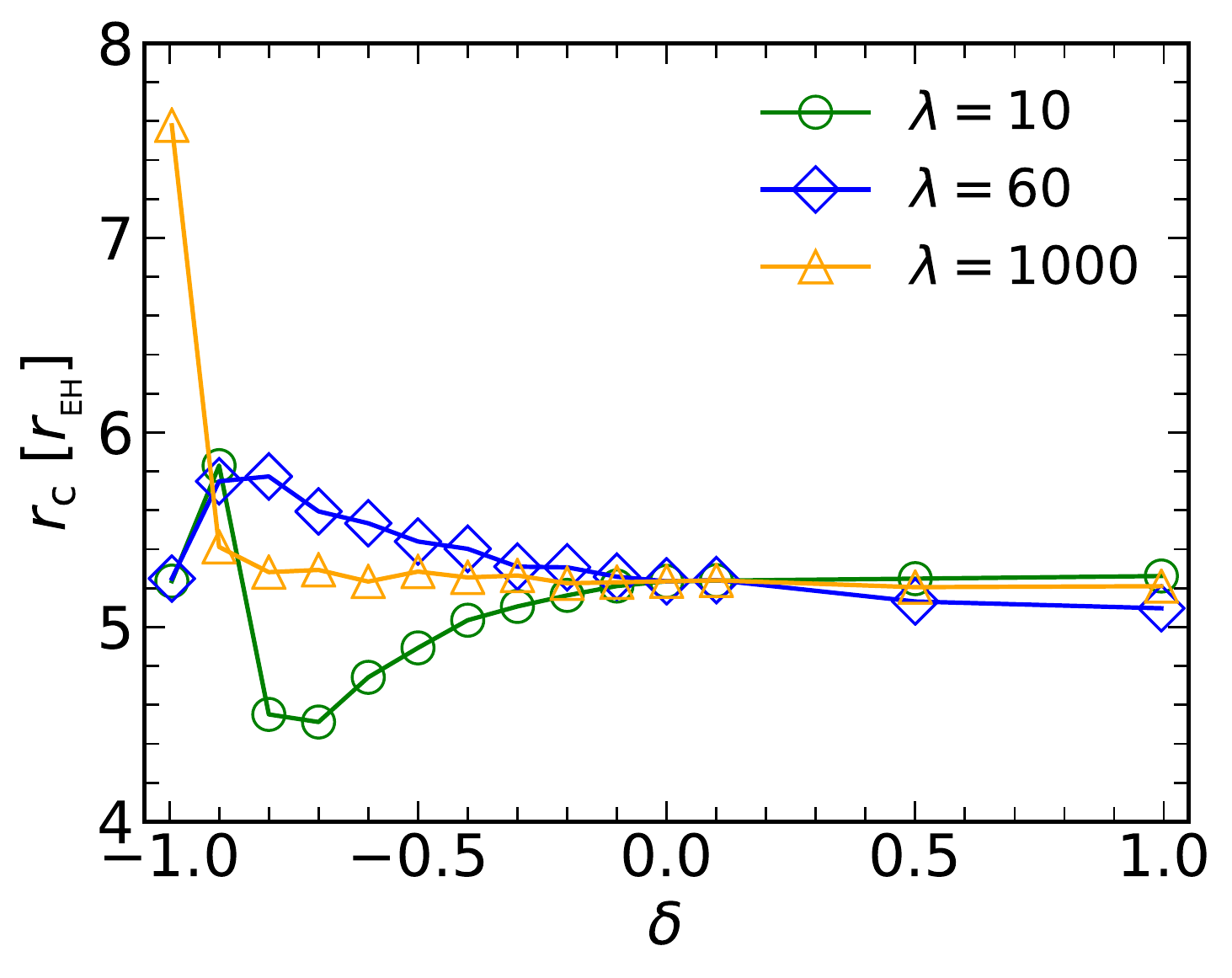}\hspace{-0.1cm}\\
\vspace{-0.55cm}
\hspace{-0.42cm}
\includegraphics[width=1\columnwidth]{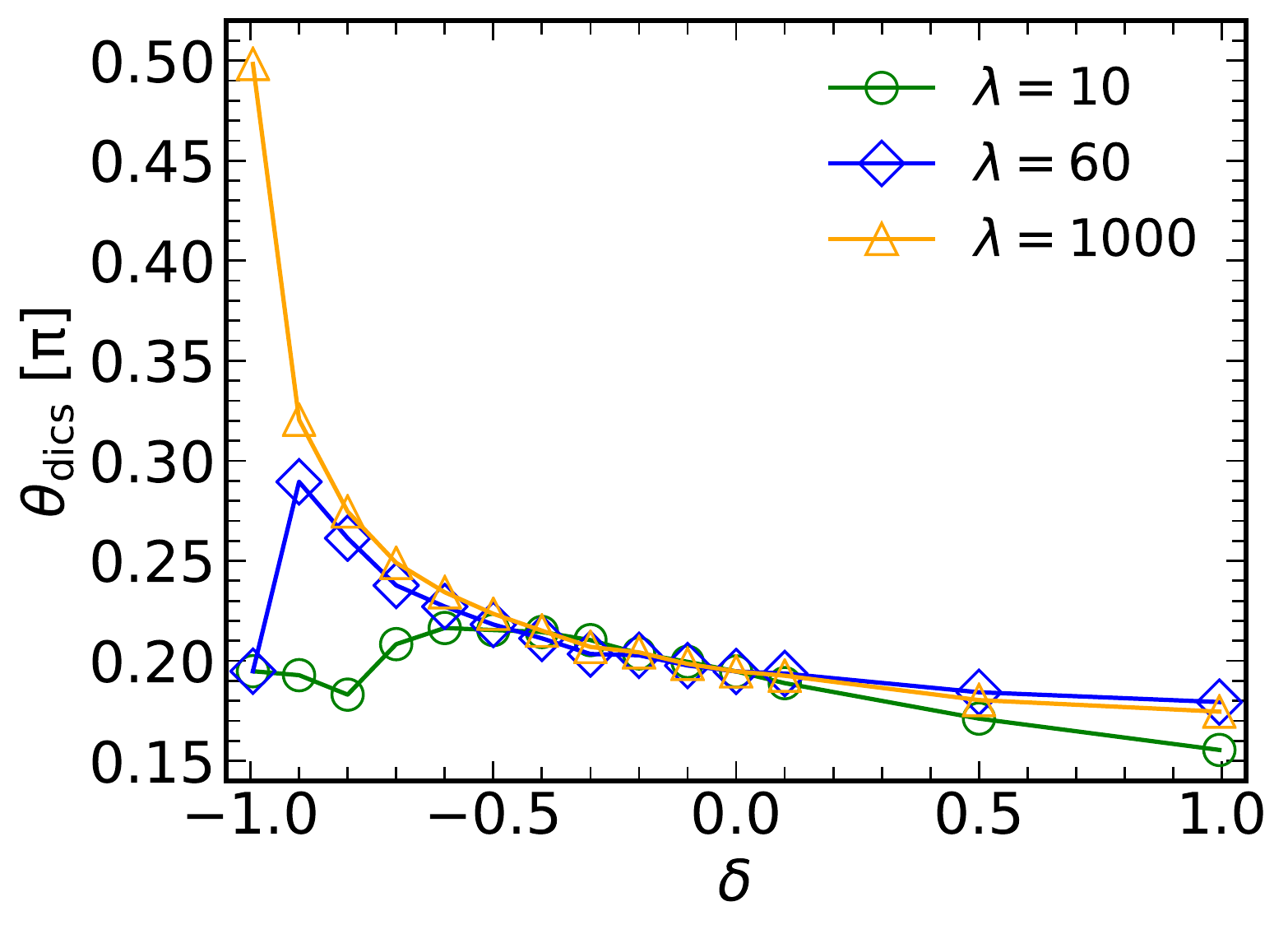}\hspace{0.1cm}
\caption{Dependence on $\delta$ of the position of the center of the disc $r_{\rm c}$ in units of  $r_{\rm {}_{EH}}$ (top) and of the angular thickness of the disc measured with respect to the $x$-axis at $r_{\rm c}$ (bottom). Note that the dependence of both  quantities on $\delta$ does not change with the magnetic field strength.}
\label{fig:rc_theta}
\end{figure}

Fig.~\ref{fig:2dmorphology} shows that the inner edge of each torus is at the same distance of the YBH horizon since we are using as unit of distance the event horizon size. The same occurs for the location of the cusp of the potential, as confirmed by the values reported in Table \ref{tab:params}.  Nevertheless, we notice differences in the location of the center of the discs $r_{\rm c}$ and of their outer edge $r_{\rm out}$. 
The top plot in Fig.~\ref{fig:rc_theta} displays the position of the center of the tori in units of the event horizon radii of the corresponding YBH as function of  $\delta$ and for all three values of $\lambda$. Since the center of the disc is defined as the location of the minimum of the potential, it does not depend on the magnetic field strength.  We observe noticeable changes in the position of the disc center for negative values of $\delta$. For the case $\lambda=60$, $r_{\rm c}$ increases monotonically with $r_{\rm EH}$ as $\delta\rightarrow -1$, up to $\delta=-0.8$ where $r_{\rm in}\approx 5.8r_{\rm EH}$. For $\delta=-0.995$, $r_{\rm c}$ decreases to about $5.2r_{\rm EH}$. Note, however, that in terms of the YBH mass, the radial position of the center of the disc always increases monotonically (see Table \ref{tab:params}).

The bottom plot of Figure~\ref{fig:rc_theta} shows the corresponding angular thickness - expressed in radians - between the surface of the tori, defined by the equipotential surface, and the $x$-axis as a function of $\delta$. In the Schwarzschild case, $r_{\rm c} \sim 5.3\ r_{\rm EH}$ and the angular thickness is $\theta_{\rm disc}\sim \pi/5$. Therefore, for negative values of $\delta$ more elongated discs with higher angular thickness are obtained, as shown in the bottom  plot of Fig.~\ref{fig:rc_theta}.

The dependence we have just described is affected by the value of the length scale parameter $\lambda$. The corresponding results for $\lambda=10$ and 1000 are also plotted in Fig.~\ref{fig:rc_theta}. For $\lambda=10$,  $r_{\rm c}$ decreases monotonically with $r_{\rm EH}$ as $\delta\rightarrow -1$, up to $\delta\approx-0.8$. For even more negative values of $\delta$, $r_{\rm c}$ increases. The discs are in general smaller than in the $\lambda=60$ case and their angular sizes, which depend weakly with $\delta$, are also smaller. Finally, for $\lambda=1000$ both $r_{\rm c}$ and $\theta_{\rm disc}$ increase monotonically with $r_{\rm EH}$ as $\delta\rightarrow -1$. The largest discs with the highest angular thicknesss are found for this value of $\lambda$. It is relevant to note that for $\lambda = 0$, both the top and the bottom panels of Fig.~\ref{fig:rc_theta} would show flat curves, since in that case, the modification induced by the parameter $\delta$ would only act as a correction of the mass parameter $M \rightarrow M/(\delta + 1)$ that cannot induce any change in the morphology of the disk. However, when $\lambda \neq 0$, it can be seen that we can define an effective mass function $M(r)$ such as 
\begin{equation}
M(r) = \frac{M(\delta e^{-r/\lambda} + 1)}{\delta + 1}\,.
\end{equation}
Then, the effective mass $M(r)$ seen by the disk is not a constant, and is different from the asymptotic mass as seen by an observer at $r \rightarrow \infty$ which is $M(r\rightarrow \infty) = M/(\delta + 1)$. These two facts are the reason of the deviations from the morphology expected for a Schwarzschild BH.

\begin{figure*}
\centering
\includegraphics[width=0.68\columnwidth,height=0.57\columnwidth]{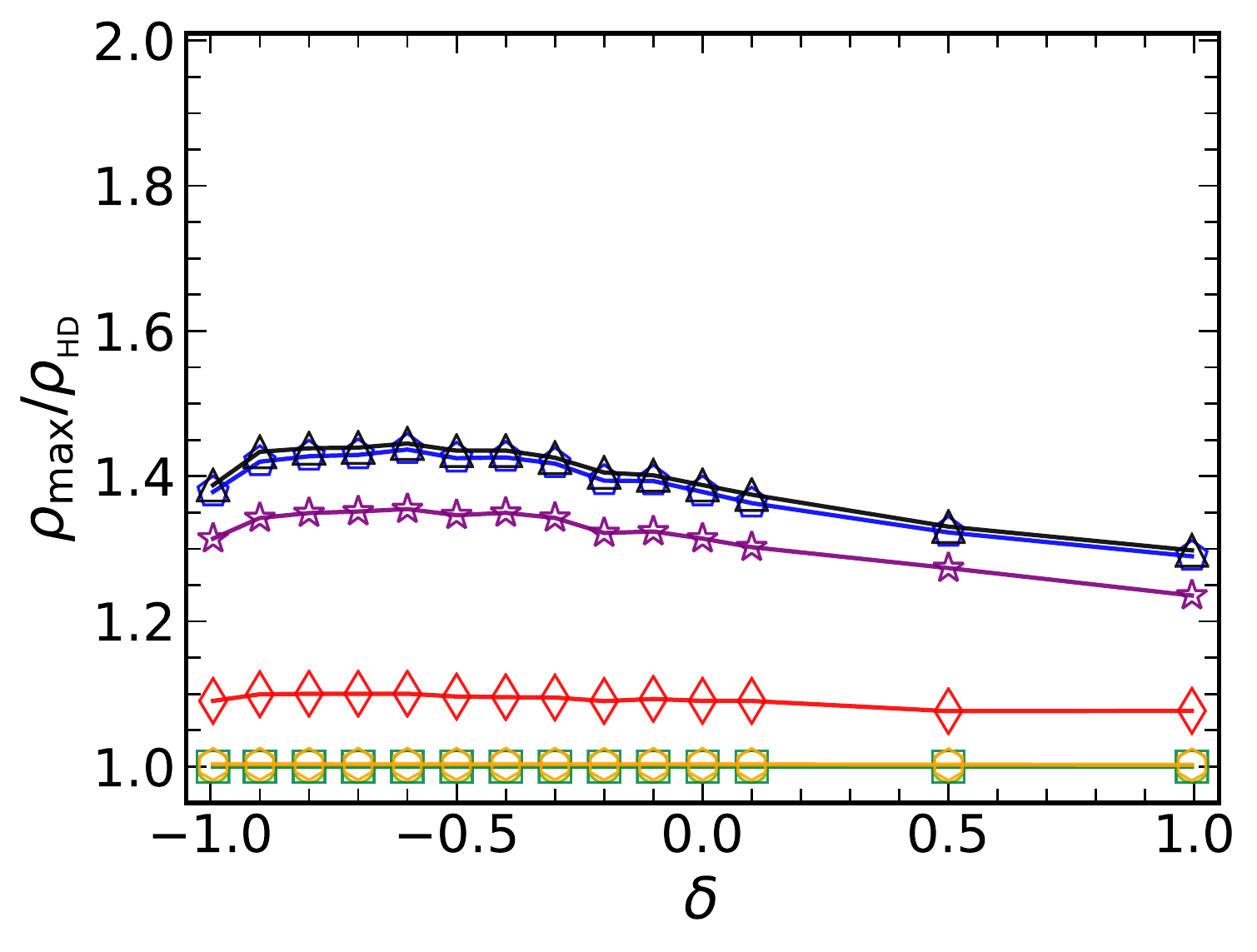}\hspace{-0.2cm}
\includegraphics[width=0.68\columnwidth,height=0.57\columnwidth]{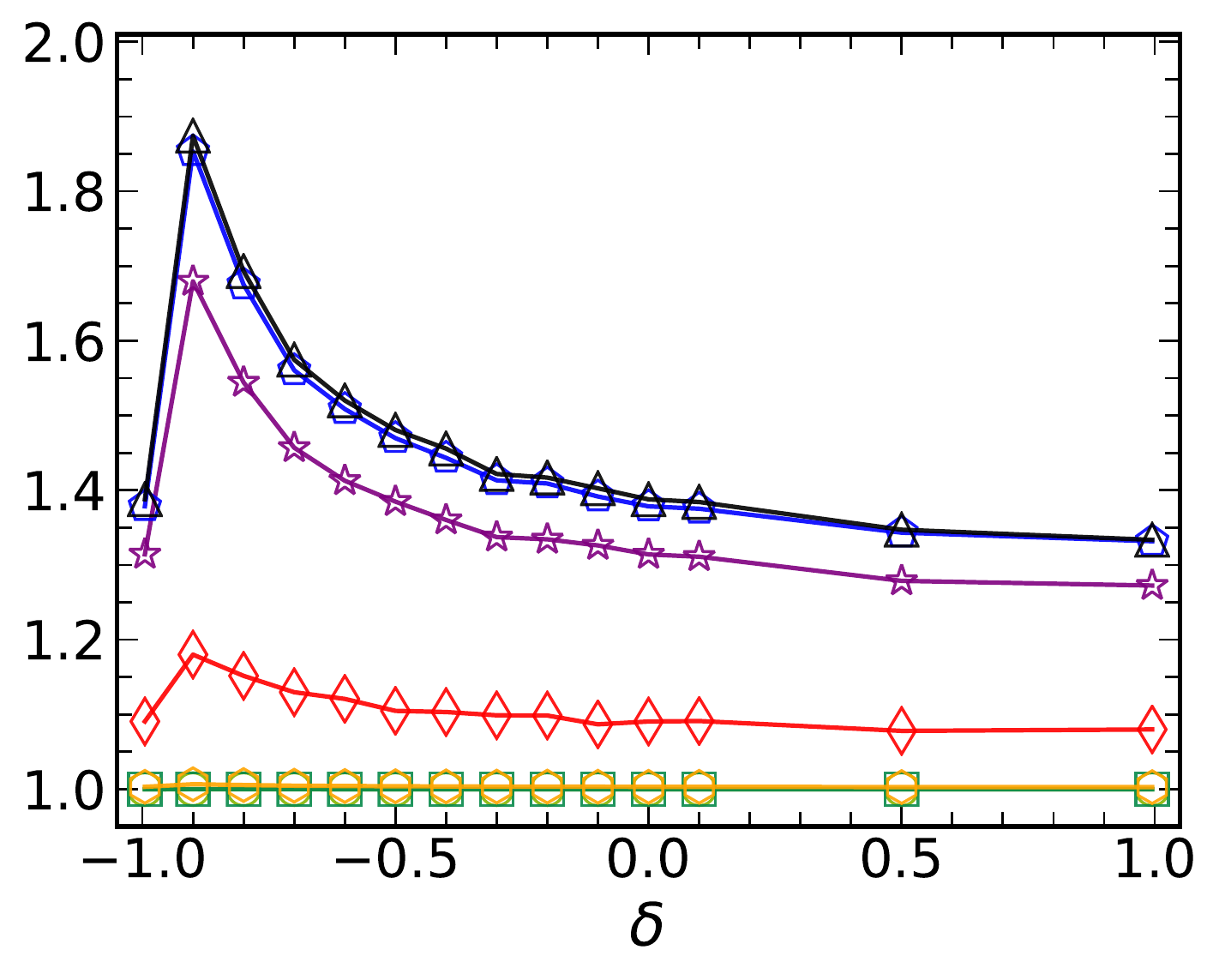} \hspace{-0.2cm}
\includegraphics[width=0.68\columnwidth,height=0.57\columnwidth]{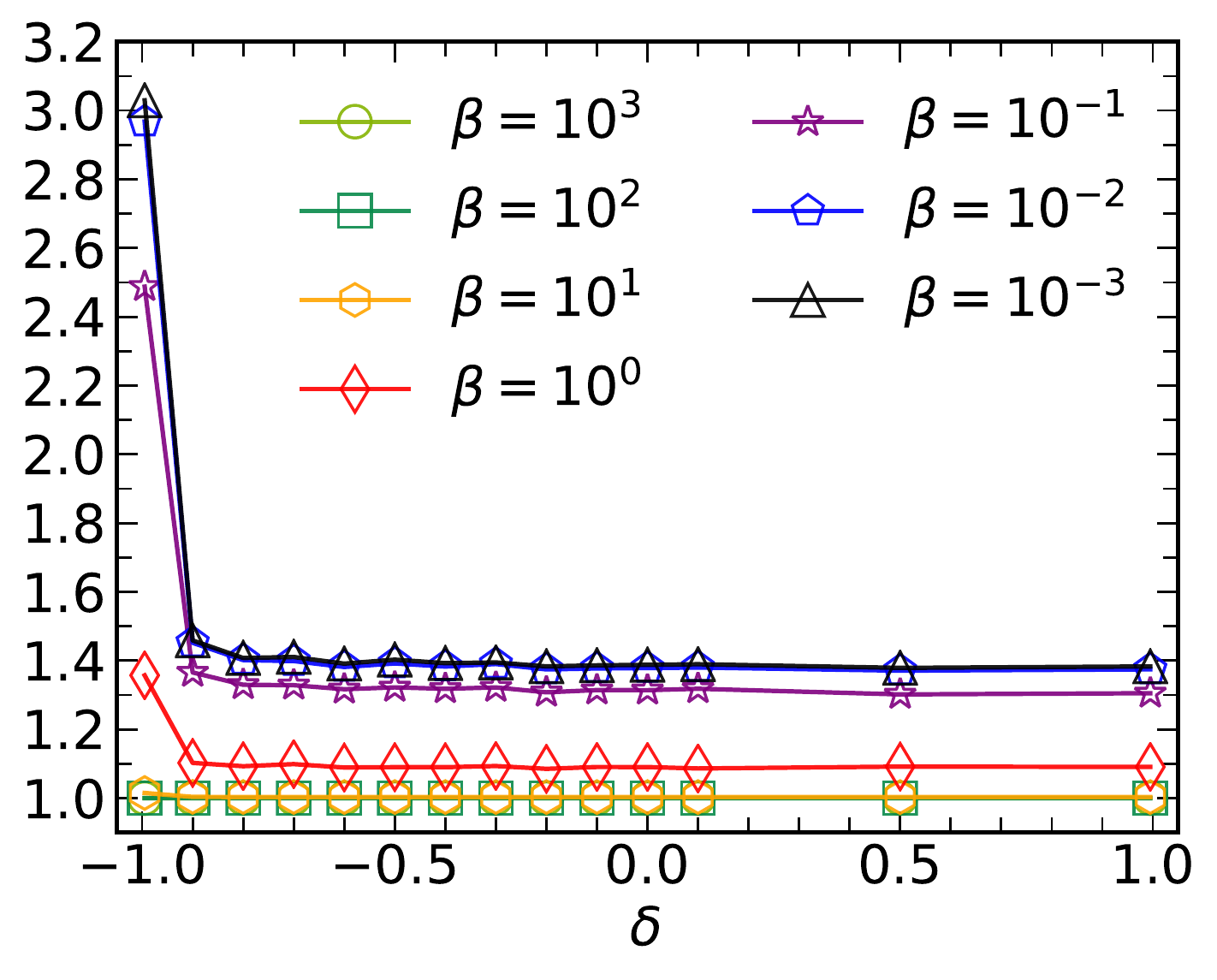}
\caption{Maximum of the rest-mass density normalized by the corresponding value for a nonmagnetized disc $\rho_{\rm {}_{HD}}$ (see Table \ref{tab:params}), for $\lambda=10$ (left),  $\lambda=60$ (middle), and  $\lambda=10^{3}$ (right). The maximum density increases with the magnetization, in agreement with the Schwarzschild black hole case. The largest increments are found for $\delta<0$.} 
\label{fig:max}
\end{figure*}

The range of variation of $r_{\rm out}$ with $\delta$ in our models is also significant, as reported in Table \ref{tab:params}. For $\lambda=10$, $r_{\rm out}\sim 21-70\ r_{\rm EH}$ , for  $\lambda=60$, $r_{\rm out}\sim 35-114\ r_{\rm EH}$ and for $\lambda=1000$, $r_{\rm out}\sim 47-1100\ r_{\rm EH}$. This effect is a consequence of the particular value of the gravitational-centrifugal potential gap for each model (which is a nonlinear function of the Yukawa-like potential), which increases as we move from positive to negative values of $\delta$, modifying the Roche lobes of each YBH solutions and the morphology and thermodynamics of the equilibrium torus. Such behavior depends directly on the Yukawa-like potential; for values of $\lambda$ comparable with the radial extent of the disk the exponential function in Eq.~\eqref{eq:Yukawa} has an important contribution on the space-time. On the other hand, for large values of $\lambda$ the exponential part goes to zero and the potential mostly depends on $1/(\delta + 1)$, consistent with the geometry and the thermodynamics of the magnetized discs (see Figs. \ref{fig:rc_theta}, \ref{fig:max}, \ref{fig:masses}).

\begin{figure*}
\centering
\includegraphics[width=0.68\columnwidth,height=0.64\columnwidth]{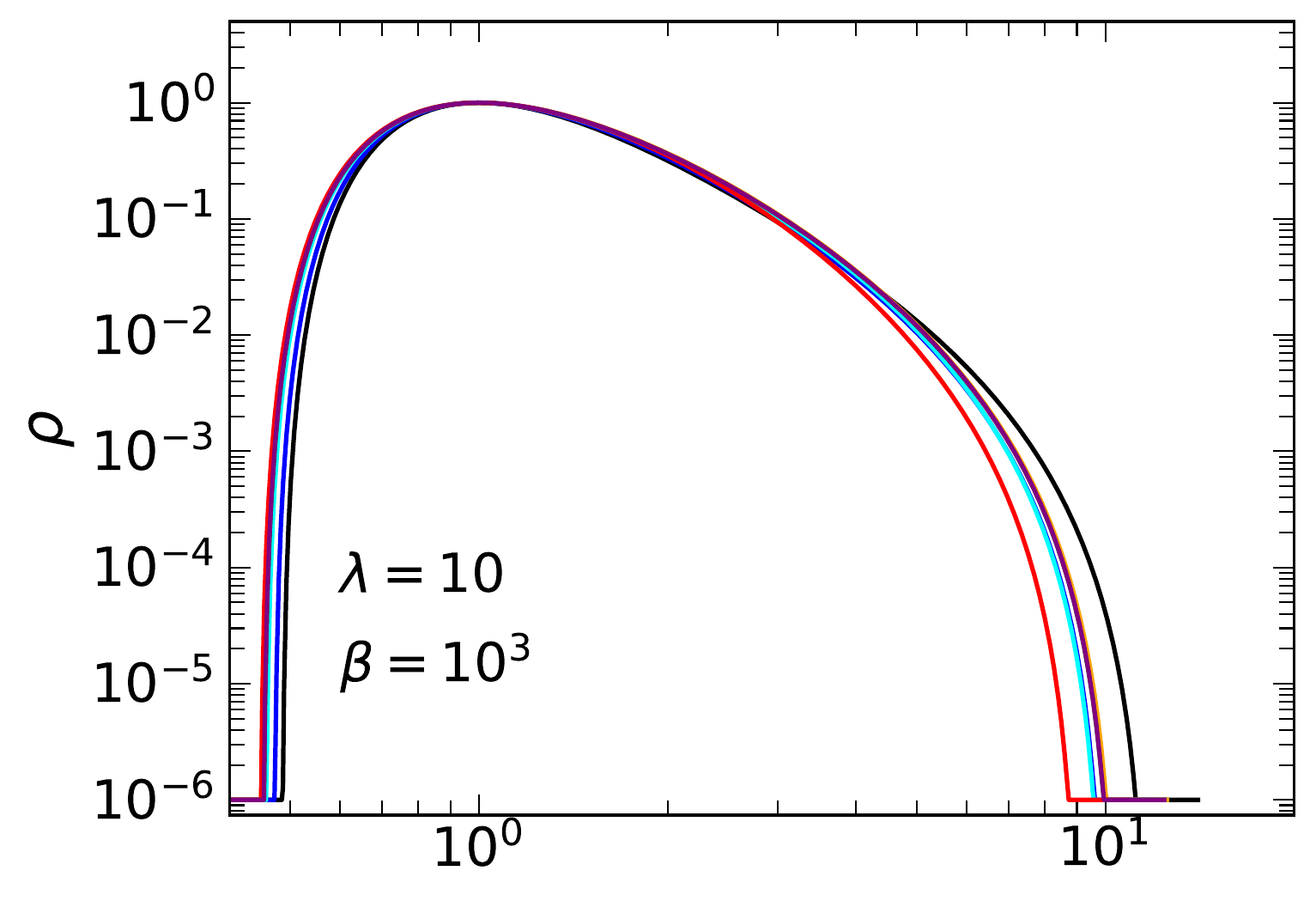}\hspace{-0.1cm}
\includegraphics[width=0.68\columnwidth,height=0.64\columnwidth]{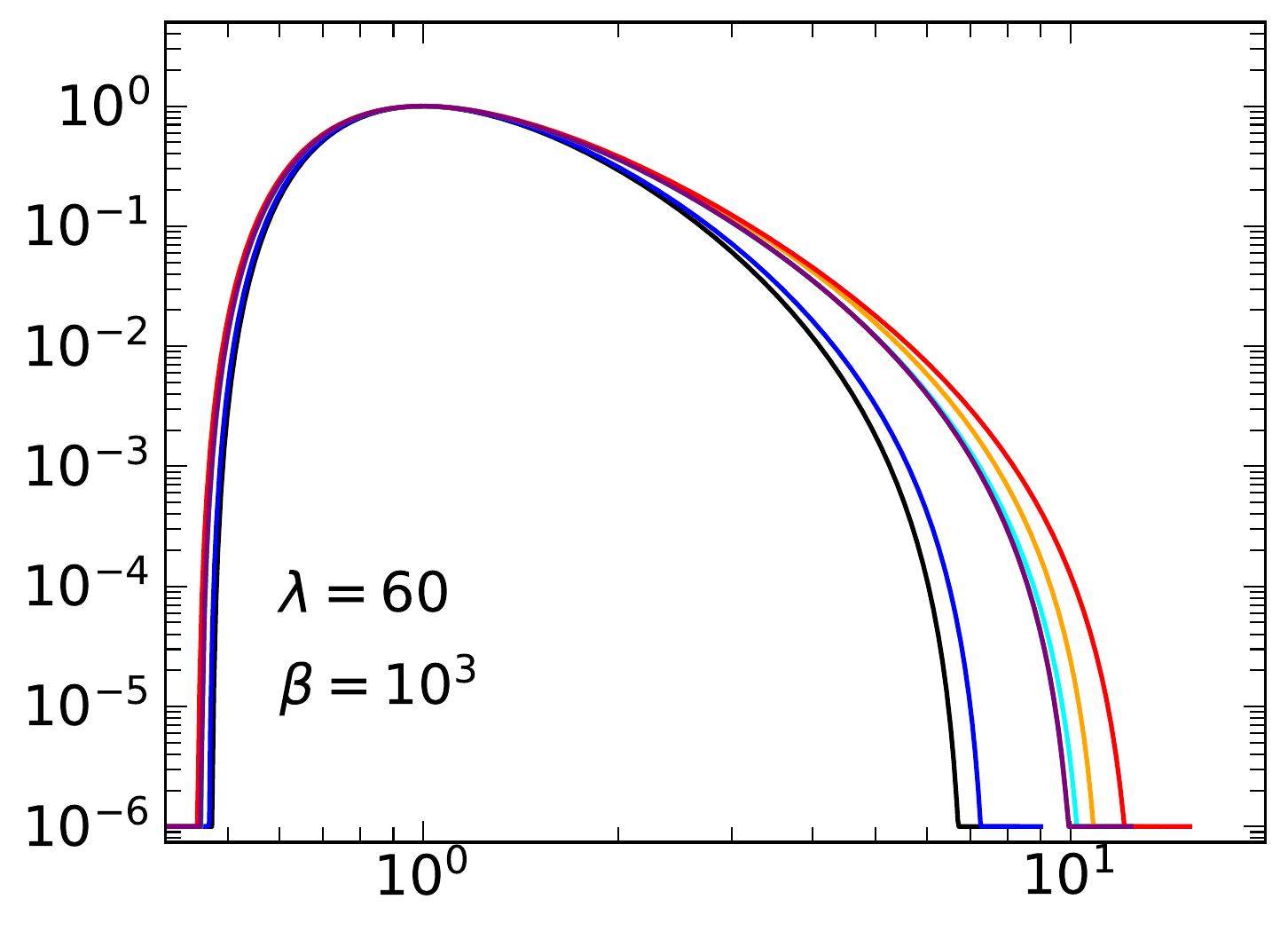}\hspace{-0.1cm}
\includegraphics[width=0.68\columnwidth,height=0.64\columnwidth]{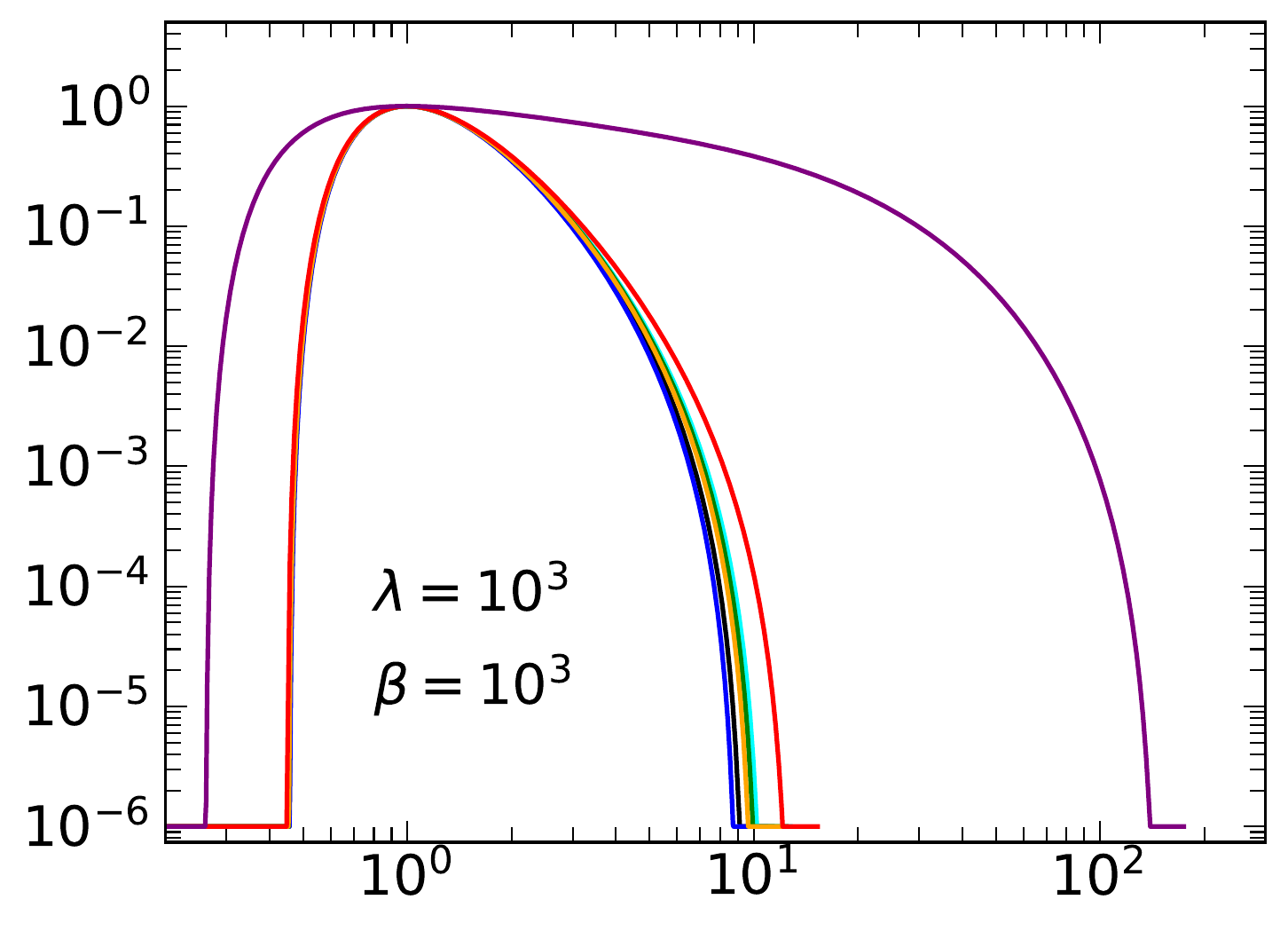}\\
\vspace{-0.2cm}
\includegraphics[width=0.68\columnwidth,height=0.64\columnwidth]{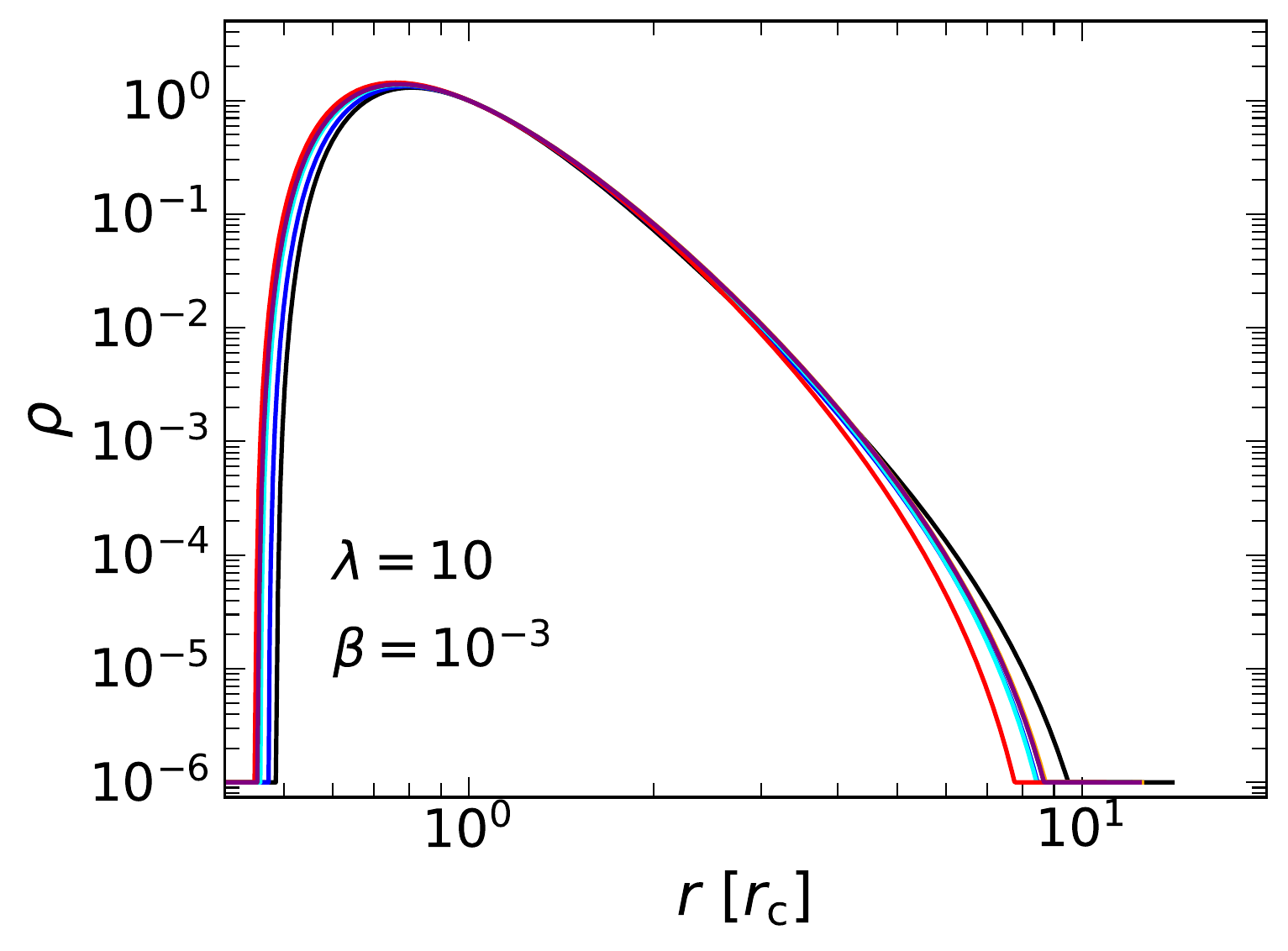} \hspace{-0.15cm}
\includegraphics[width=0.68\columnwidth,height=0.64\columnwidth]{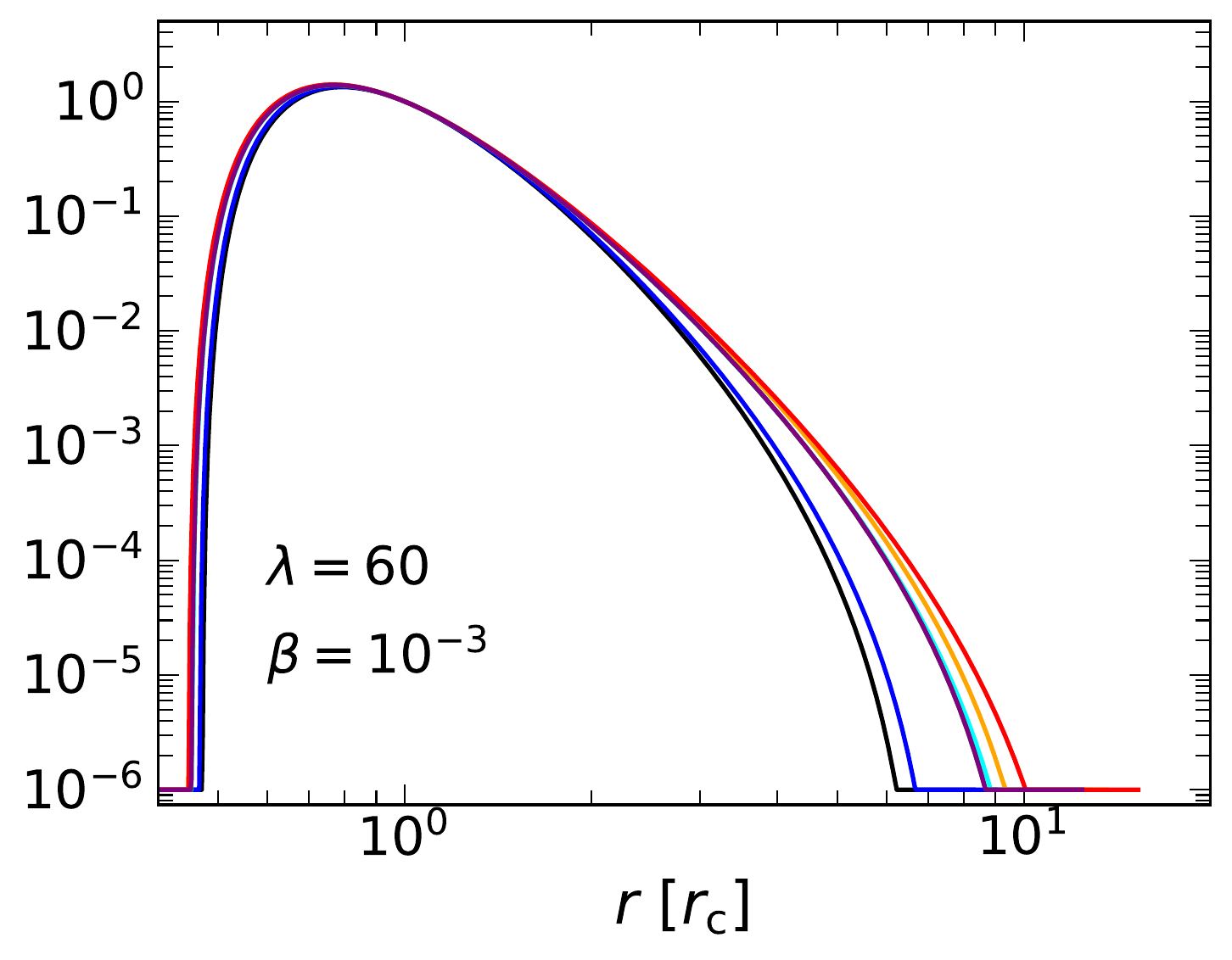} \hspace{-0.15cm}
\includegraphics[width=0.68\columnwidth,height=0.64\columnwidth]{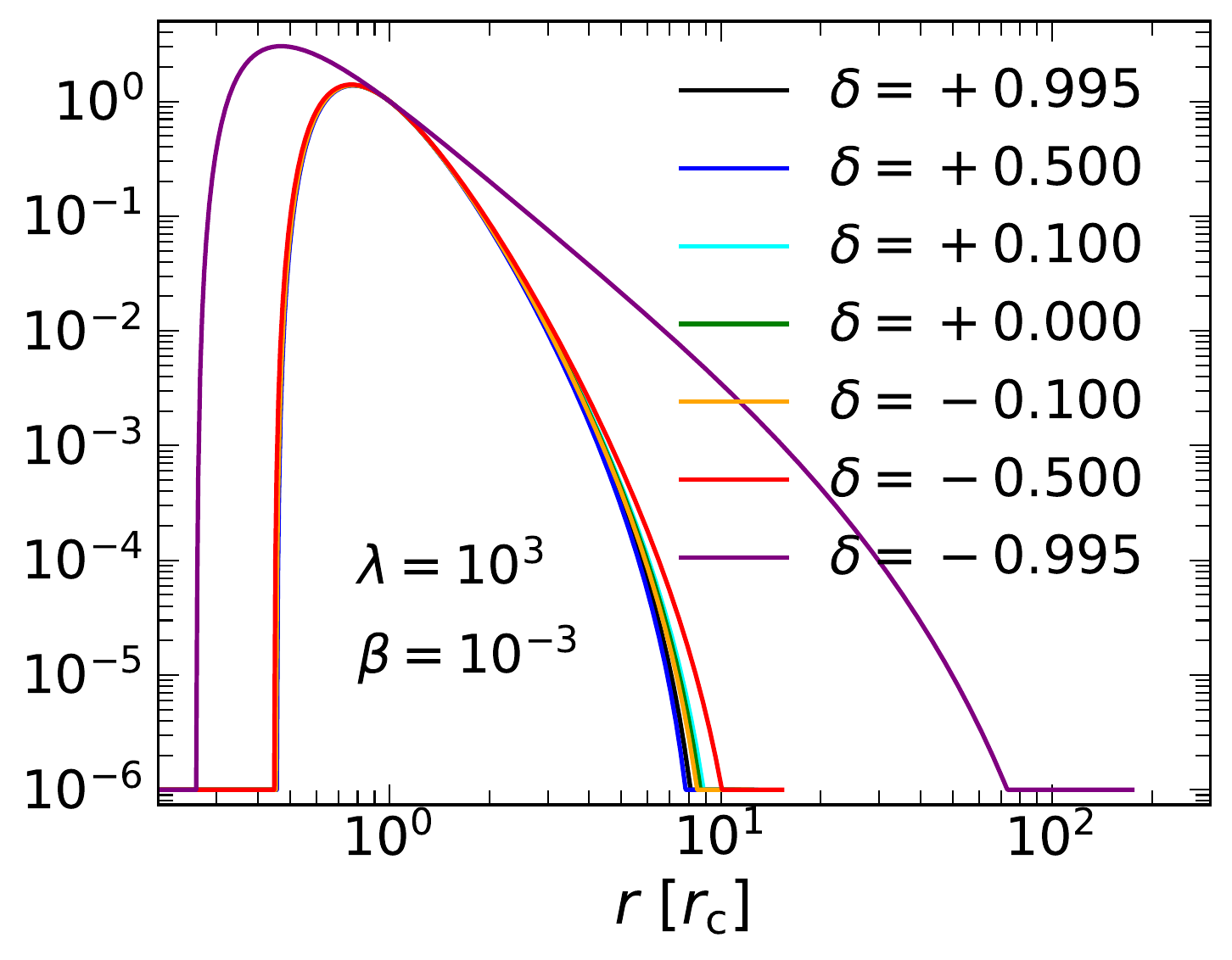}
\vspace{-0.2cm}
\caption{Radial profiles of the rest-mass density of the magnetized discs in logarithmic scale at the equatorial plane. The left panels correspond to the YBH space-time with $\lambda=10$, the middle panels to $\lambda=60$ and the right panels to  $\lambda=10^{3}$. The top panels show low magnetized discs and  the bottom ones display highly  magnetized cases. In each plot seven values for $\delta$ are shown. The radial coordinate is in units of the center of the disc to facilitate the comparison.}
\label{fig:1Dr_Density}
\end{figure*}


\subsection{Tori thermodynamics}

The differences found in the geometry of the discs are accompanied by quantitative differences in the physical magnitudes characterizing the matter content of our models and their thermodynamics. Figure \ref{fig:1Dr_Density} shows radial profiles along the equatorial plane ($\theta=\pi/2$) of the rest-mass density in logarithmic scale for two values of the magnetization parameter and for all values of parameters $\lambda$ and $\delta$. The profiles are shown for both an unmagnetized disc ($\beta=10^{3}$) and a highly magnetized one ($\beta=10^{-3}$). For YBH space-time, the maximum of the rest-mass density increases for high magnetized discs and its location shifts toward the inner edge of the torus. This is in agreement with what is found for the case of Schwarzschild space-time~\citep{Gimeno-Soler:2019}. However, the deviations are not large and they only become significant for $\delta=-0.995$. For some of the models, the form of the YBH space-time allows us to build discs which are almost filled by high-density regions (e.g.~model with $\lambda=10^3,\ \delta=-0.995$ and $\beta=10^3$).

\begin{figure*}
\centering
\includegraphics[width=0.68\columnwidth,height=0.57\columnwidth]{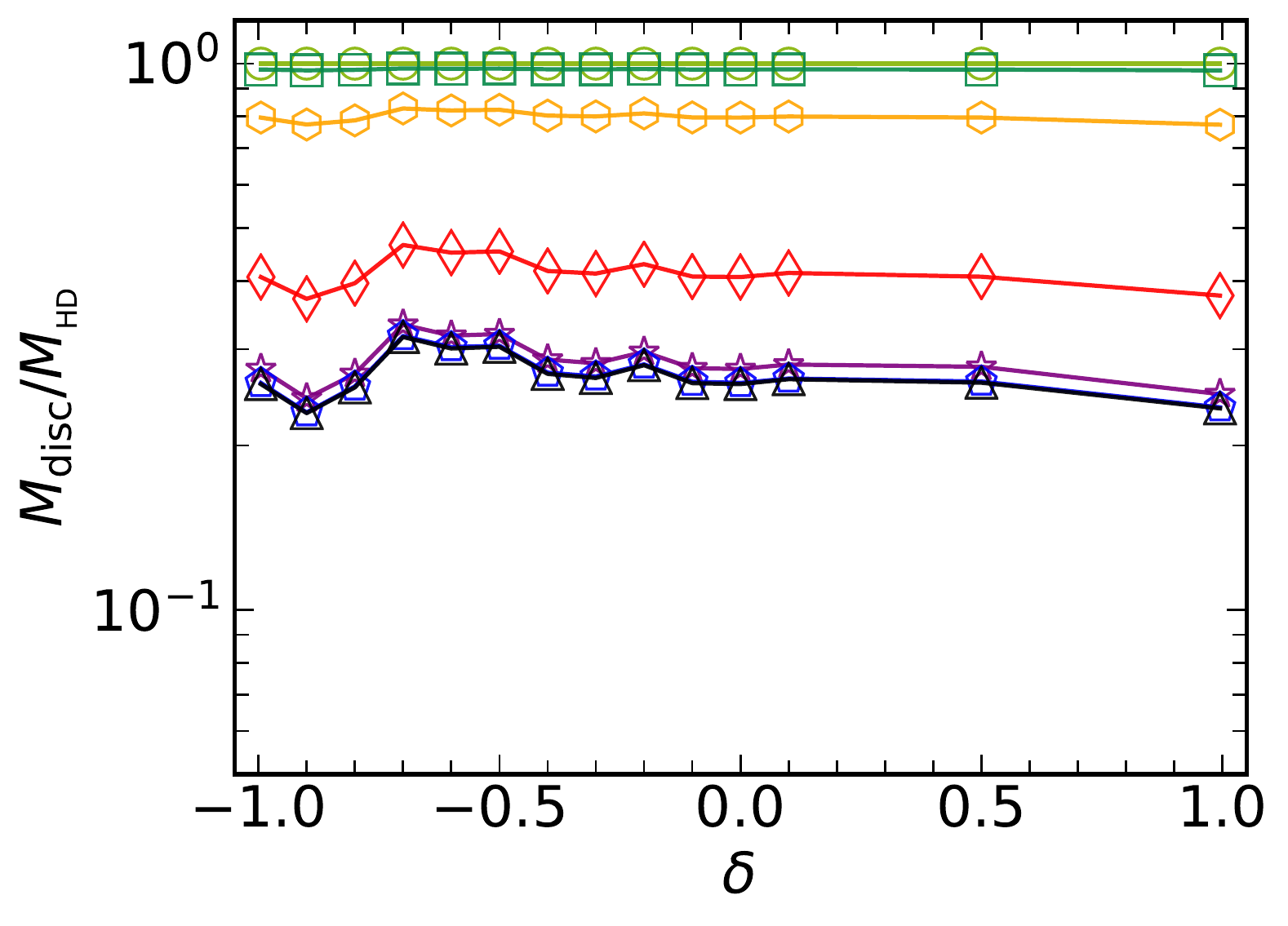}\hspace{-0.2cm}
\includegraphics[width=0.68\columnwidth,height=0.57\columnwidth]{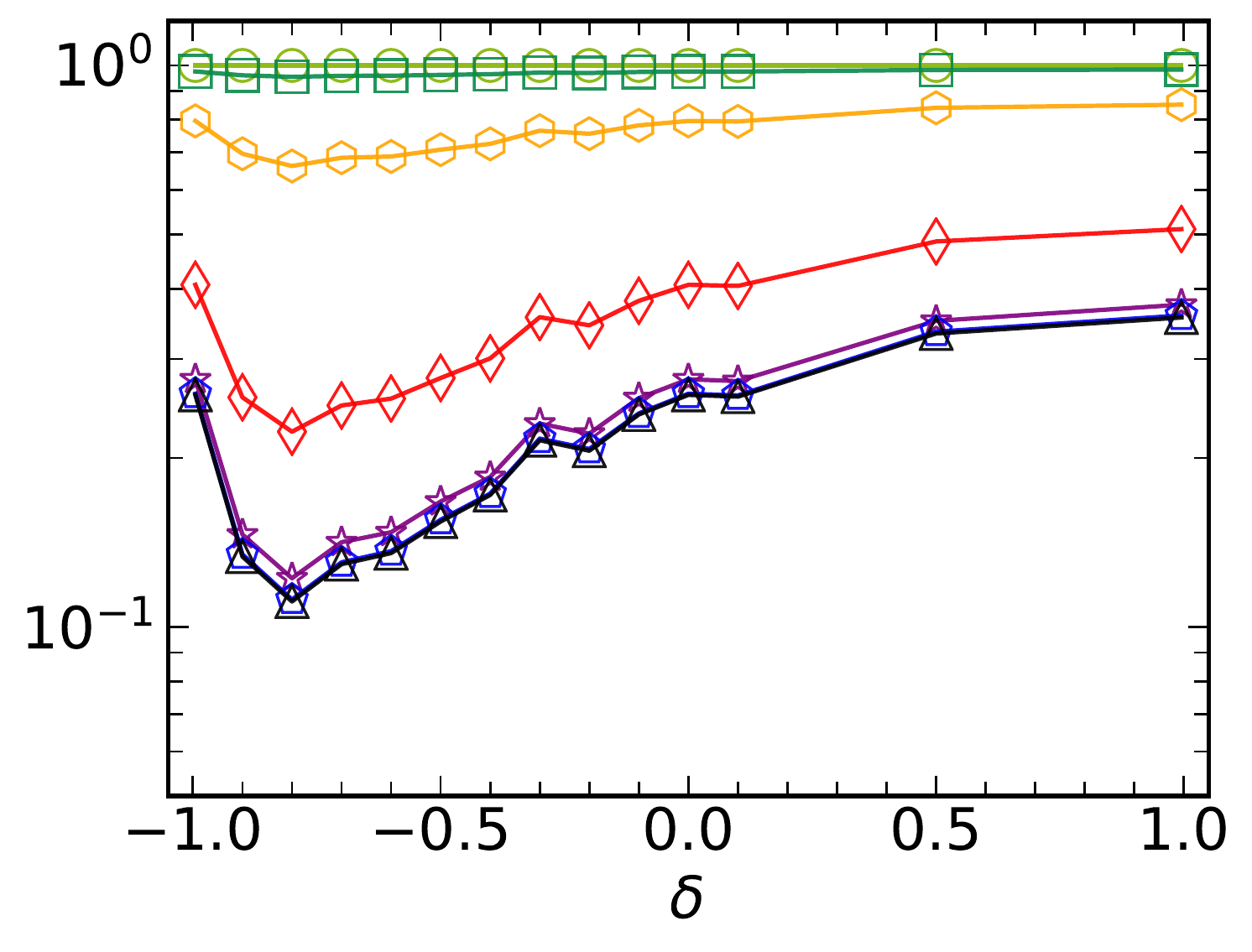} \hspace{-0.3cm}
\includegraphics[width=0.68\columnwidth,height=0.57\columnwidth]{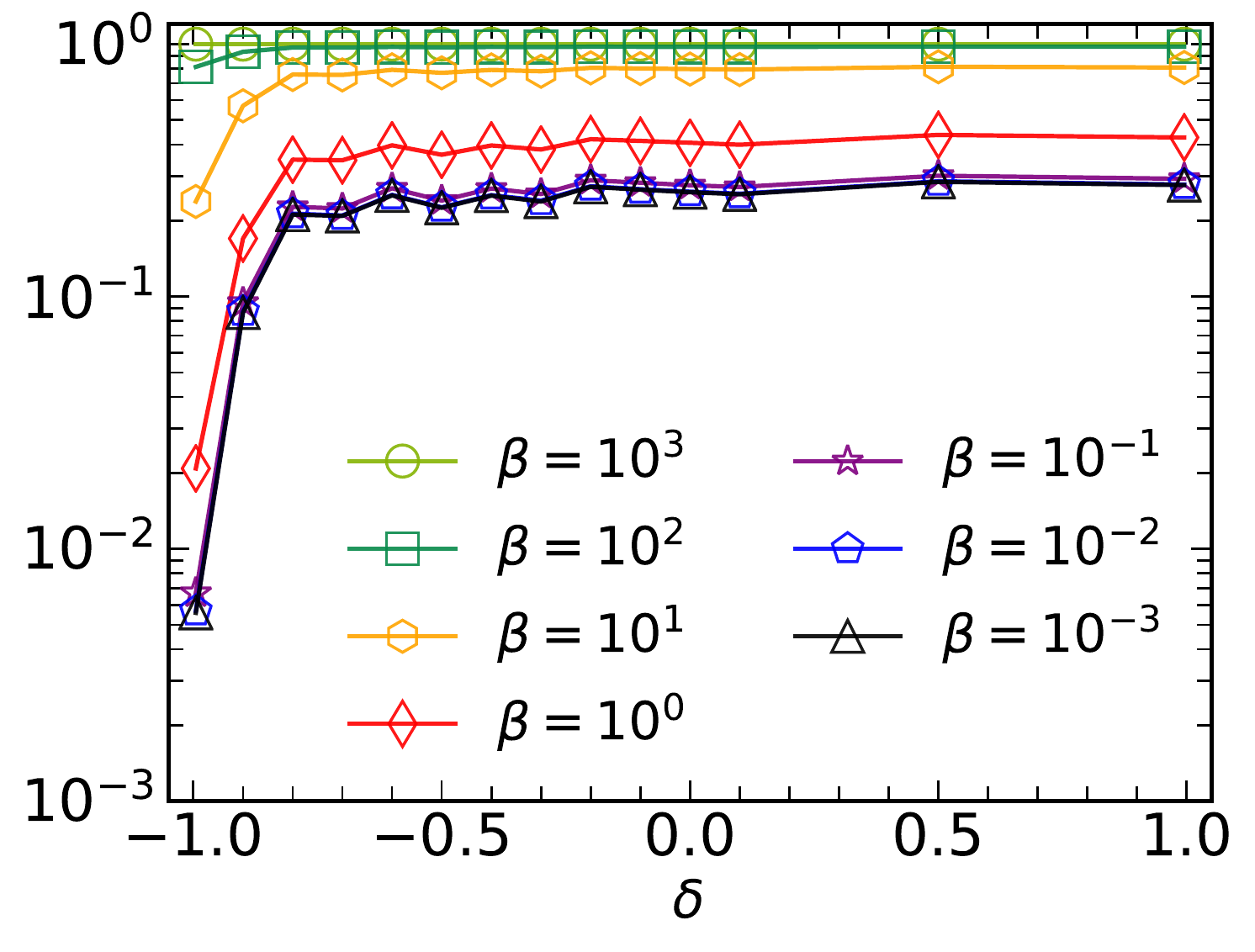}\\ 
\caption{Disc mass as a function of $\delta$  normalized by the mass of an unmagnetized disc, $M_{\rm {}_{HD}}$ (see Table \ref{tab:rho_and_m}), for different magnetizations. From left to right the panels correspond to 
$\lambda=10$,  $\lambda=60$, and $\lambda=10^{3}$, respectively. The disc mass decreases when the magnitude of the magnetic field increases. The least massive torus is obtained for $\lambda=10^{3}$ and $\delta \to -1$.} 
\label{fig:masses}
\end{figure*}

Table \ref{tab:rho_and_m} reports, for all of our models, the values of the maximum of the rest-mass density and of the baryon mass of the disc, defined as
\begin{eqnarray}
M_{\rm disc}=\int \sqrt{\gamma}W \rho d^{3}x\,,
\end{eqnarray}
where $W$ is the Lorentz factor. This table allows to quantify the effects on these two quantities of the  parameters $\delta$ and $\lambda$ that characterize the YBH spectime and to find the dependence on the magnetic-field strength, from unmagnetized to highly magnetized discs. We assume that the mass of the disc is $10\%$ of the mass of the black hole for the unmagnetized case, $M_{\rm HD}=0.1M$, fixing the rest-mass density, $\rho_{\rm HD}$. The highest value of $M_{\rm disc}$ is attained for the unmagnetized model ($\beta=10^3$) and the mass of the tori decreases monotonically as the magnetization increases.

Figures \ref{fig:max} and \ref{fig:masses} show the maximum of the rest-mass density and of the baryon mass of the disc as a function of $\delta$, normalized by  $\rho_{\rm HD}$ and $M_{\rm HD}$, respectively. Both quantities do not show a strong dependence on $\delta$, irrespective of the value of $\lambda$, except for values of $\delta$ close to -1, where the largest deviations are found.
The increase in $\rho_{\rm max}$ is monotonic with the increase of the disc magnetization, and it is similar for all three values of $\lambda$.
Except for $\delta\rightarrow -1$ the maximum value of the density is fairly constant and about $1.4\ \rho_{\rm HD}$ which is the value found for tori around a Schwarzschild black hole by \cite{Gimeno-Soler:2019}. Only when $\delta\rightarrow -1$ higher values of the rest-mass density are found, namely $\sim 1.54, 1.9, 3.0 \ \rho_{\rm HD}$ for $\lambda=10, 60, 1000$ respectively. In those cases, the total baryon mass of the discs decreases to about $\sim M_{\rm HD}/3,\ M_{\rm HD}/10$ and $M_{\rm HD}/100$ when the strength of the magnetic field increase.

\begin{figure*}
\centering
\includegraphics[width=0.61\columnwidth,height=0.321\columnwidth]{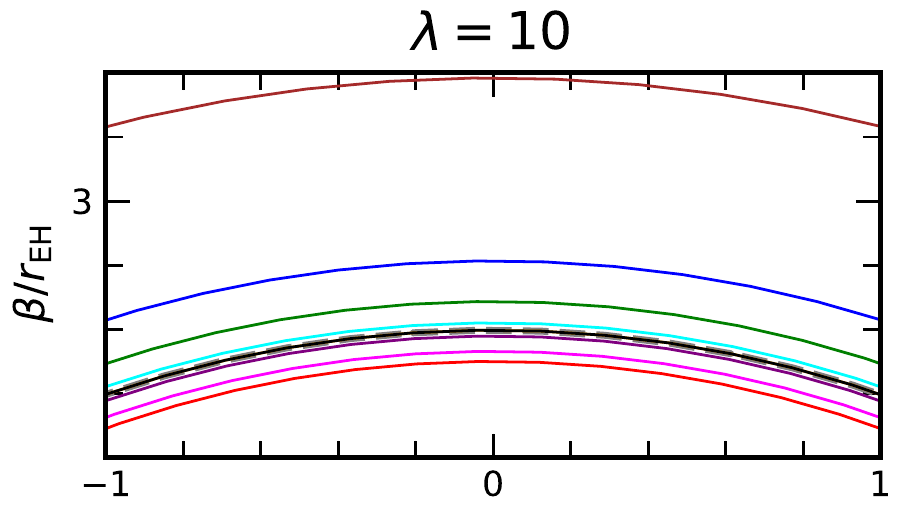}\hspace{-0.1cm}
\includegraphics[width=0.61\columnwidth,height=0.321\columnwidth]{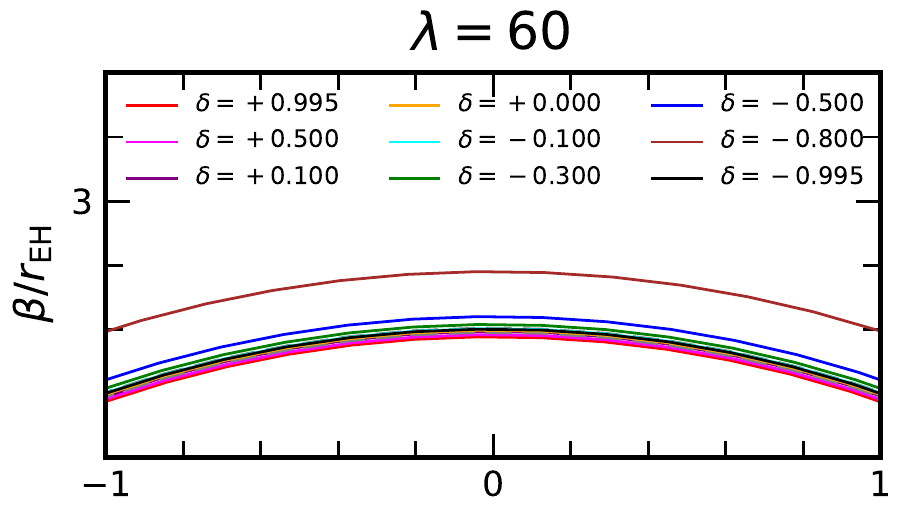} \hspace{-0.2cm}
\includegraphics[width=0.61\columnwidth,height=0.321\columnwidth]{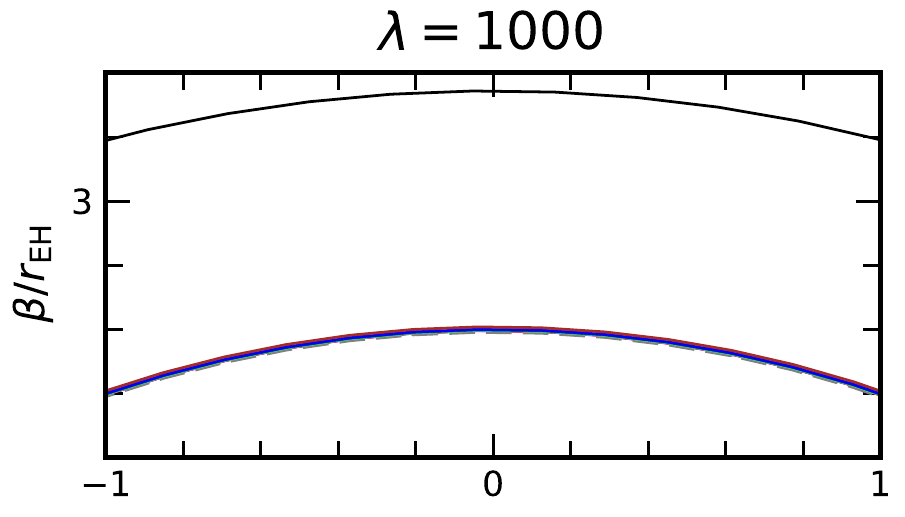}\\
\includegraphics[width=0.61\columnwidth,height=0.61\columnwidth]{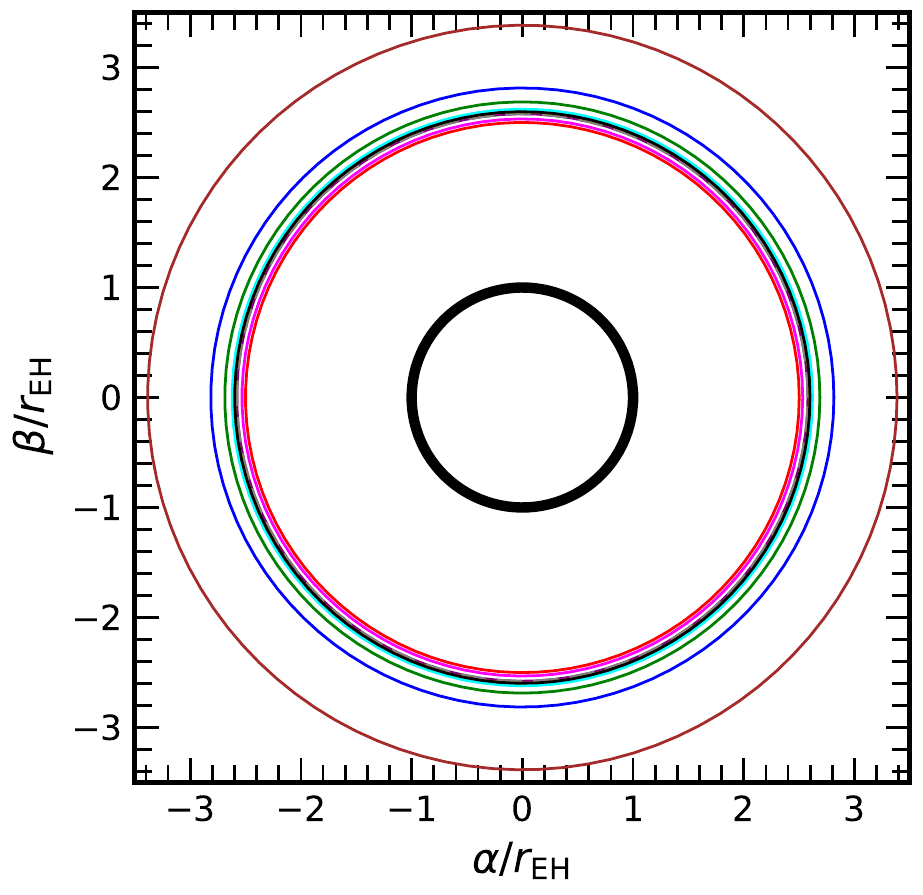}\hspace{-0.1cm}
\includegraphics[width=0.61\columnwidth,height=0.61\columnwidth]{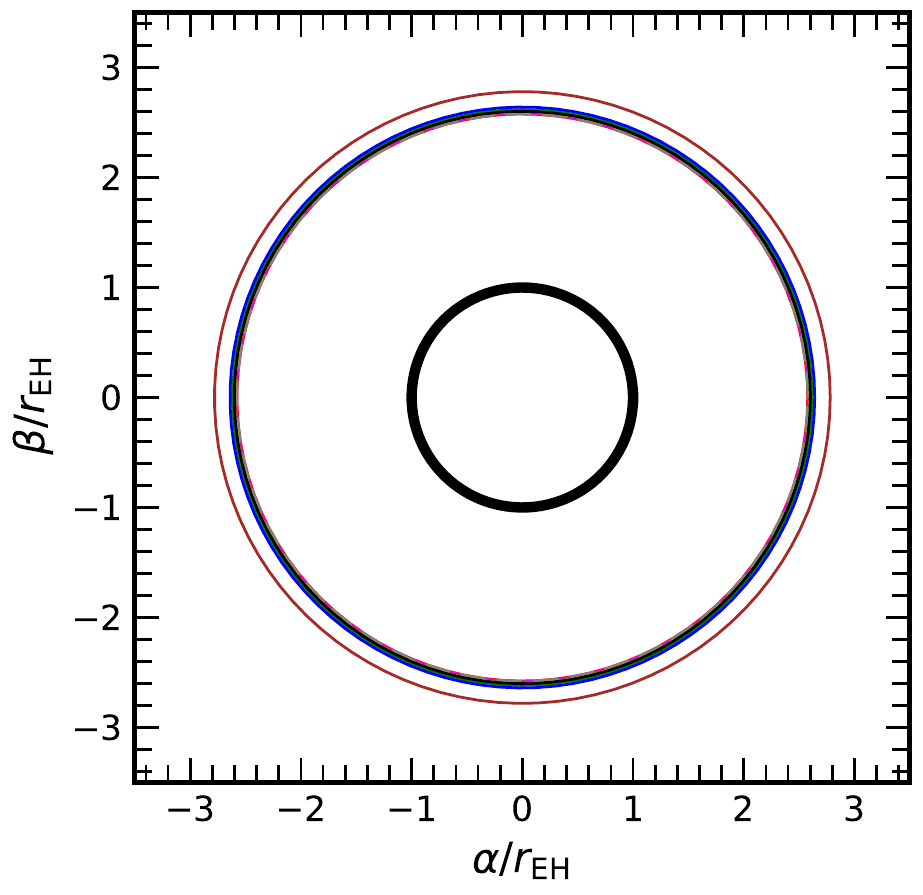} \hspace{-0.1cm}
\includegraphics[width=0.61\columnwidth,height=0.61\columnwidth]{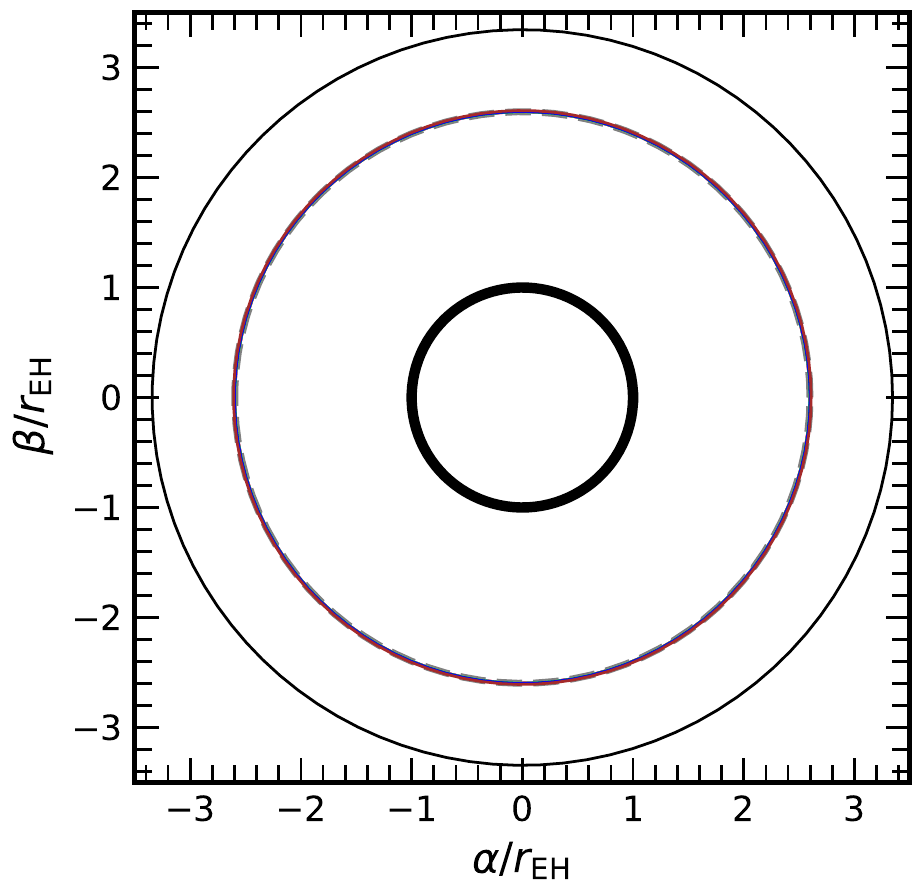}\\
\includegraphics[width=0.61\columnwidth,height=0.321\columnwidth]{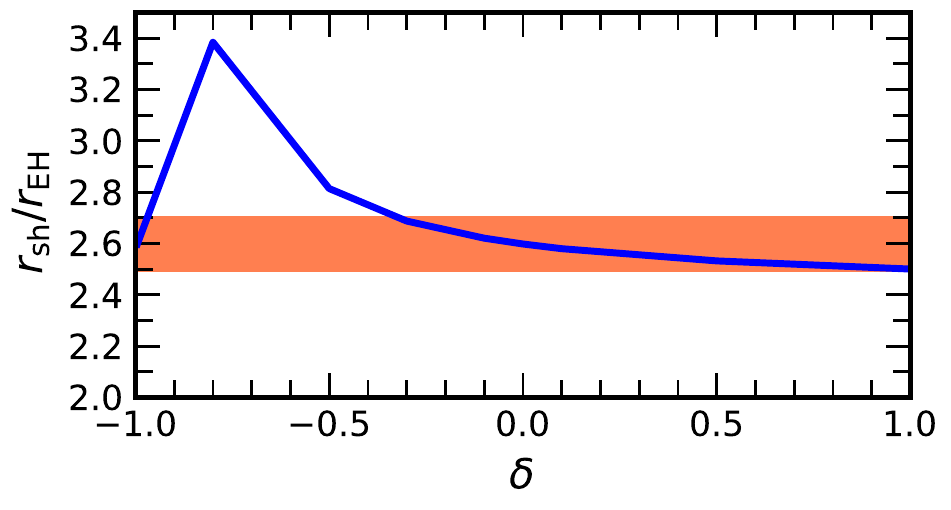}\hspace{-0.1cm}
\includegraphics[width=0.61\columnwidth,height=0.321\columnwidth]{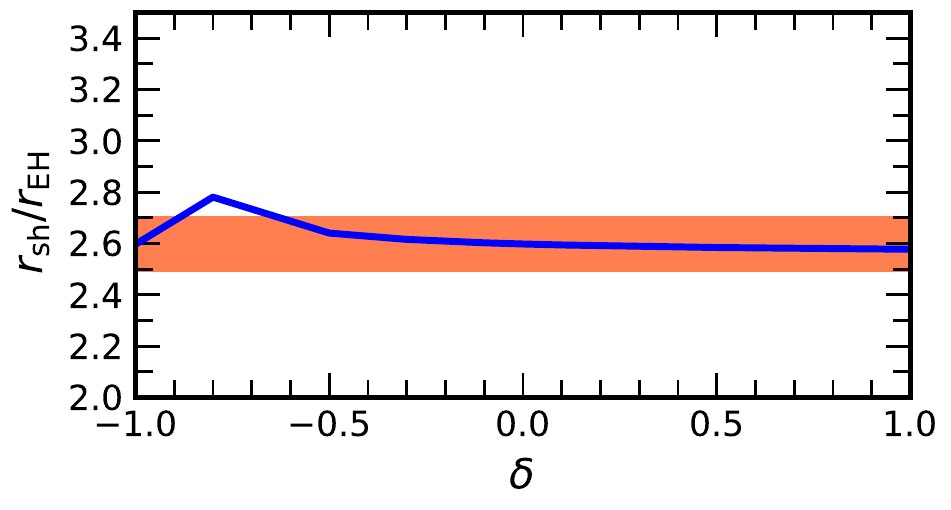} \hspace{-0.1cm}
\includegraphics[width=0.61\columnwidth,height=0.321\columnwidth]{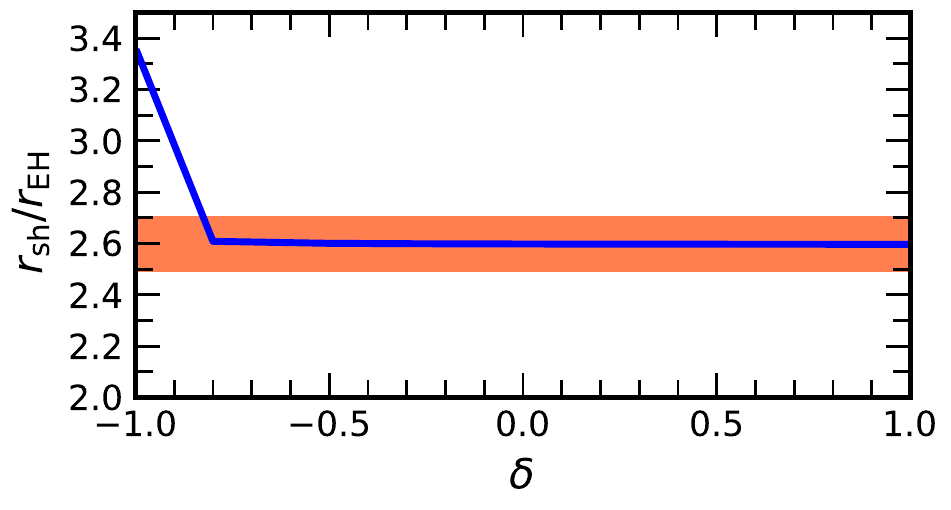}
\caption{Photon ring size (shadow size) of different YBH space-times with varying $\delta$ parameter and length scales $\lambda$. The full rings are displayed in celestial coordinates $(\alpha,\beta)$ normalized by the event horizon size in the middle panels, while a closeup is shown in the top panels. The shadow radius as a function of $\delta$ is shown in the bottom panels (blue curves), where the orange regions indicate the range of the shadow size for maximally rotating Kerr black holes.
The dashed gray lines correspond to the Schwarzschild black hole shadow radius, $\rm r_{sh}=3\sqrt{3} r_{g}$~\citep{Johannsen2010} and the black line is the radius of the event horizon. An observer view angle $\rm i^{o}=\pi/2$ at infinity is assumed.}
\label{fig:photonring}
\end{figure*}

\subsection{Constraining the YBH parameters with the photon ring size}

The Event Horizon Telescope observations of the black hole shadow in M87~\citep{EHT_M87_PaperI,EHT_M87_PaperVI} and the forthcoming 
observations of SgrA* provide a laboratory to test general relativity and modified theories of gravity by using the shadow properties. In particular, the fundamental property M87's shadow revealed in EHT observations is the photon ring. This is the geometrical region in which space-time bends light in such a way that photons must follow circular orbits. For an observer at infinity, the photon ring defines the apparent size of the shadow, i.e.~the shadow size, that only depends on the gravity of space-time. We turn now to analyze how the shadow size of our YBH solution differs from that of a Schwarzschild black hole, constraining the parameters of the space-time with the size of the photon ring. To this aim we solve the geodesic equations for photon (null) trajectories in the YBH space-time given by Eq.~\eqref{eq:Yukawa} following the procedure developed by \citet{Hioki2009, Hou2018}. We neglect any contribution from  the emissivities and absorptivities of the magnetized torus (i.e.~we do not consider the spectral content of the light) because for our stationary models we assume that the emission from the photon ring is brighter than in the disc, but the disc is sufficiently illuminated so that a photon ring is visible. A detailed assessment of the shadows based on numerical evolutions of our YBH-torus systems will be presented elsewhere. The apparent shape of the YBH shadow is computed by introducing the celestial coordinates $(\alpha, \beta)$ assuming a source located at infinity  and a viewing angle $\rm i^{o}=\pi/2$. Such coordinates  measure apparent angular distances of the image on the celestial sphere. The shadow of the black hole is defined by a bright ring at the radius of the lensed photon sphere or photon ring~\citep{EHT_M87_PaperIV}. For a Kerr black hole at different viewing angles such radius varies between $\rm r_{sh} \sim 3\sqrt{3}r_{g} \pm 4\%$ \citep{Johannsen2010}, where $3\sqrt{3}r_{g}$ is the photon ring of the Schwarzschild black hole, and $r_{g}$ is the gravitational radius. The $\pm 4\%$ limits in the variation of the size of the Schwarzschild black hole shadow correspond to the maximally rotating Kerr black hole cases (with Kerr dimensionless spin parameter $a=\pm 1$). 

Figure \ref{fig:photonring} shows the shadow size at a viewing angle $\rm i^{o}=\pi/2$ for the YBH space-times listed in  Table~\ref{tab:params}.
Each column corresponds to one of our three length scales $\lambda=10, 60$, and $10^3$. In the middle panels we show the complete shape of the photon rings in celestial coordinates, normalized by the event horizon of the black hole, for  representative values of the $\delta$ parameter. In the top panels we show a close-up of the upper region of the ring while the bottom panels display the radius of the photon ring $r_{\rm sh}$ as a function of $\delta$. The orange regions in these plots are the shadow sizes estimated from general relativity for Kerr black holes. Almost all of our models are consistent with Event Horizon Telescope observations~\citep{EHT_M87_PaperIV}. The largest deviations are obtained for the most negative values of $\delta$, irrespective of $\lambda$, albeit for $\lambda=60$ the variations are the smallest. On the other hand, for positive values of $\delta$ the photon ring size only increases slowly with $\delta$ for the two lowest values of $\lambda$ we consider.

\begin{figure}
\centering
\includegraphics[width=0.95\columnwidth,height=0.95\columnwidth]{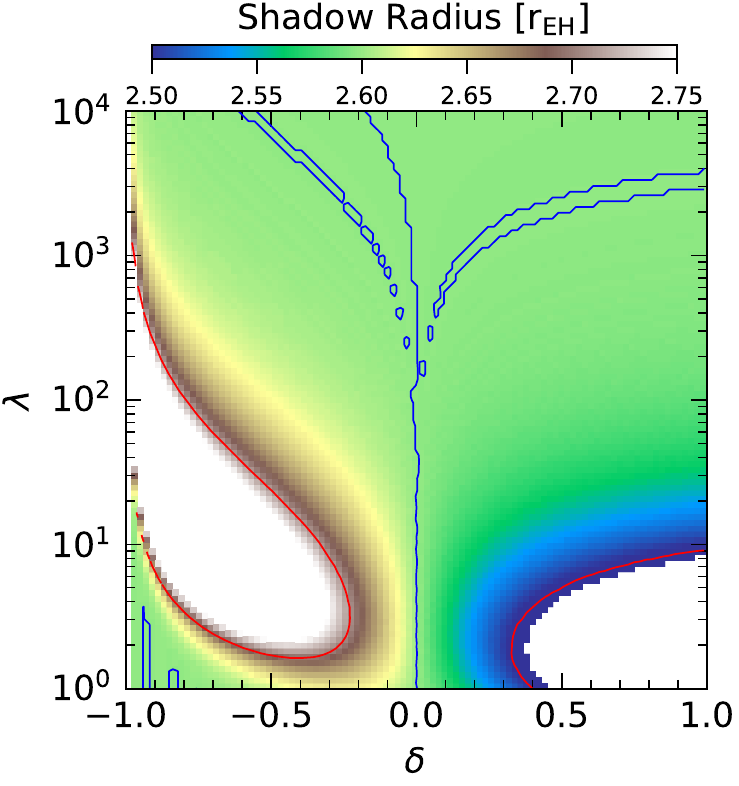}
\caption{Shadow radius of the YBH as function of $\delta$ and $\lambda$, normalized by the event horizon. The blue contour corresponds to shadow size for Schwarzschild black hole $\rm r_{sh} \sim 3\sqrt{3}$, in $r_{g}$ radius units and red contours to $\rm r_{sh} \pm 4\%$.}
\label{fig:all}
\end{figure}

To obtain a better understanding of the dependence of the shadow size on $\delta$ and $\lambda$ we finish our analysis by exploring
$10^{4}$ space-time models varying both parameters. The results are displayed in Fig.~\ref{fig:all} which shows the shadow radius normalized by the event horizon radius as a function of $\delta$ and $\lambda$. The blue isocontour corresponds to a family of YBH with the same photon ring size as the Schwarzschild black hole. The red iscontours are the bounds for $3\sqrt{3} r_{g} \pm 4\%$. All YBH photon rings between these two isocontours are consistent with general relativity Kerr black hole solutions. Photon rings of YBH which are not allowed in this range can be neglected (white regions in the figure). Figure \ref{fig:all} shows that YBH with $\delta>0$ and $\lambda<50$ generate small photon rings, whereas large photon rings are produced for $\delta<0$ for all values of $\lambda$. Irrespective of the sign of $\delta$ we observe nonlinear correlations between the shadow size and the $\delta$ parameter. In particular, at small scales, for $\lambda < 20$, the effects of the Yukawa-like gravitational potential are strong, leading to a high variability of the photon ring size - from minimum to maximum - when the $\delta$ parameter decreases from $\delta \sim 0.3$ to $\delta\sim -0.25$. This is expected since light bending is more noticeable in stronger gravity regimes. Correspondingly, for asymptotic values $\lambda\rightarrow \infty$ the YBH photon rings tend asymptotically to the Schwarzschild photon ring. 

Applying the $\delta$ parameter constriction to the supermassive black holes of M87 and the galactic center ($\lambda\sim 60$) we find that  astrophysically accepted values for $\delta$ would fall in the range $-0.75< \delta < 1.0$. For our stationary accretion disc solutions this implies that we can neglect very thick torus with large angular thickness, $\theta_{\rm disc}>0.26 \pi$, for which the location of the center is $\rm r_{c}>6r_{\rm {}_{EH}}$ and have large event horizon, $r_{\rm {}_{EH}}\sim 400M$. The maximum densities in YBH-torus systems compatible with the constrained range of $\delta$ are in agreement with the values found in general relativity~\citep{Gimeno-Soler:2019} with small deviations $\sim \pm 0.2$.

\section{Summary}
\label{sec:Sum}

We have presented stationary solutions of geometrically thick discs (or tori) with constant angular momentum and endowed with a self-consistent toroidal magnetic field distribution surrounding a nonrotating black hole in $f(R)$-gravity. The particular $f(R)$-gravity model we have employed introduces a Yukawa-like modification to the Newtonian potential, encoded in a single parameter $\delta$ and whose specific values affect the disc configurations compared to the general relativistic case. We have built models for different magnetic field strengths, from low magnetized discs (essentially hydrodynamic) to highly magnetized tori. This has been achieved by adjusting the  magnetization parameter $\beta$, \ie the ratio of thermal pressure to magnetic pressure, in the range $\rm log_{10}\beta \in \{-3,3\}$. 
Our stationary solutions have been obtained numerically, employing the approach discussed in detail in \cite{Gimeno-Soler:2019}.

The characteristics of our solutions have been quantified by analyzing the central density of the discs, their baryonic mass, their geometrical size and angular thickness, as well as the effects of the deviations of the YBH metric from the Schwarzschild metric. We have found that in the general relativistic limit ($\delta=0$) our models reproduce our previous results for a Schwarzschild black 
hole~\citep{Gimeno-Soler:2017,Gimeno-Soler:2019}. For small values of the $\delta$ parameter, corresponding to $\sim 10\%$ deviations from general relativity, we have found small geometrical variations in the models with respect to the general relativistic results, namely $\sim 2 \%$ in the event horizon size, a $\sim 5\%$ shift in the location of the inner edge and center of the  disc, and $\sim 10\%$ variation in the location of the outer edge. We note that our results are consistent with the shift in the periastron advance of about $10\%$ reported by \cite{DeLaurentis2018ahr}.

Our analysis for $|\delta|>0.1$ has revealed notable changes in the black hole solutions, particularly in the limit $\delta\rightarrow -1$. Those modifications of the gravitational potential have a large direct impact in the torus solution. We have found that the influence of the magnetic field in the disc properties  becomes stronger in this case.  In particular we have observed an increment of the YBH event horizon of about four orders of magnitude with respect to the event horizon of a Schwarzschild black hole, a $\sim 10\%$ increase of $r_{\rm c} (r_{\rm EH})$ and three orders of magnitude increase of the locations of the outer edge of the disc, $r_{\rm out}$.  

The impact of the strength of the toroidal magnetic field in the morphology of the discs follows the same trend of previous analysis for nonrotating black holes in general relativity \citep{Gimeno-Soler:2017,Gimeno-Soler:2019}. The maximum density for our most highly magnetized disc is $\sim 1.4\ \rho_{\rm HD}$ and the total mass of the disc is $\sim 0.25\ M_{\rm HD}$. The variation in these two quantities is less than $10\%$ for small deviations from general relativity. For $|\delta|>0.1$ the increment in the maximum density is a factor two with respect to general relativity and the disc mass decreases one order of magnitude. 
In addition, negative values of the YBH parameter $\delta$ influence the angular size of the discs, which become more elongated along the $z$-axis. This may affect the angular size of outflows and jets that might form when evolving these  magnetized discs in the alternative theory of gravity discussed here.

Finally, we have analyzed the differences between the YBH space-time and the Schwarzschild space-time by computing the size of the photon ring produced by a source located at infinity. This has allowed us to constrain the parameters $\delta$ and $\lambda$ of the YBH space-time when applying our approach to the supermassive black holes of M87 and SgrA$^*$. 
The variations in the photon ring reported in this work, even for small deviations from general relativity, might be measurable in upcoming observations of the galactic center by the Event Horizon Telescope Collaboration.

Even though the variations we have found in our YBH-disc models are less than $10 \%$ for realistic deviations of general relativity, namely $\delta=\pm 0.1$, nonlinear time evolution of these initial data might still disclose potential differences in observable quantities like jet power or mass accretion rates, as well as in the time-dependent morphology of the photon ring produced by a turbulent, illuminating disc. Those results will be presented elsewhere.

\section*{Acknowledgements}
We thank H\'ector Olivares, Oliver Porth, and Ziri Younsi for numerous helpful discussions and comments. ACO gratefully acknowledges support from the COST Action CA16214 ``PHAROS", the  LOEWE-Program in HIC for FAIR, and the  EU Horizon 2020 Research ERC Synergy Grant ``Black-HoleCam: Imaging the Event Horizon of Black Holes" (Grant No. 610058). JAF is supported by the Spanish Agencia Estatal de Investigaci\'on (Grant No. PGC2018-095984-B-I00) and by the Generalitat Valenciana (Grant No. PROMETEO/2019/071). M.D.L. acknowledges INFN sez. di Napoli, Iniziative Specifiche QGSKY and TEONGRAV. SM acknowledges support from DGAPA-UNAM (IN112019), Consejo Nacional de Ciencia y Teconolog\'ia (CONACyT), M\'exico (CB-2014-01 No.~240512) and CONACyT 26344. This work has further been supported by the European Union's Horizon 2020 Research and Innovation (RISE) programme H2020-MSCA-RISE-2017  Grant  No.   FunFiCO-777740. The simulations were performed on the LOEWE cluster in CSC in Frankfurt.\\

\bibliographystyle{apsrev4-1}
\bibliography{Torus}


\end{document}